\documentclass[a4paper,11pt]{article}

\pdfoutput=1 

\usepackage{fullpage}

\usepackage{mathrsfs}
\usepackage{comment}
\usepackage{centernot}
\usepackage{mathtools}
\usepackage{stmaryrd}

\usepackage{makecell}

\usepackage{latexsym,amssymb,amstext,amsmath}
\usepackage{hyperref}
\usepackage{graphbox,graphicx}
\usepackage{subcaption}
\usepackage{amsthm}
\usepackage{esint}
\usepackage{float}

\def\be {\begin{equation}}
\def\ee {\end{equation}}
\def\bea {\begin{eqnarray}}
\def\eea {\end{eqnarray}}

\theoremstyle{definition}

\usepackage{tikz}
\usetikzlibrary{positioning,arrows}
\tikzset{
block/.style={
  draw, 
  rectangle, 
  minimum height=1.5cm, 
  minimum width=4cm, align=center
  }, 
line/.style={->,>=latex'}
}

\newcommand{\A}{\mathcal{A}}
\newcommand{\D}{\mathcal{D}}
\newcommand{\F}{\mathcal{F}}

\global\dimen\footins = 6cm

\begin{document}

\begin{center}

\vspace*{ 0.0cm} {\textbf{
\LARGE
Weyl-Lewis-Papapetrou coordinates, 
\vskip 2mm
self-dual Yang-Mills equations 
and the single copy}}
\\[0pt]
\vspace{0.3cm}
\bigskip
{{\bf  Gabriel Lopes Cardoso$^{\text{a}}$}},
{{\bf  Swapna Mahapatra$^{\text{b}}$}},
{{\bf Silvia Nagy$^{\text{c}}$}}
\bigskip 

\vspace{0.cm}
{\it \small 
${}^{\text{a}}$ 
Center for Mathematical Analysis, Geometry and Dynamical Systems,
Department of Mathematics, Instituto Superior T\'ecnico, Universidade de Lisboa,
1049-001 Lisboa, Portugal\\[0.2cm]}

{\it \small 
${}^{\text{b}}$ 
Department of Physics, Utkal University, Bhubaneswar 751004, India \\[0.2cm]}

{\it \small 
${}^{\text{c}}$ 
Department of Mathematical Sciences, Durham University, Durham, DH1 3LE, UK  \\[0.4cm]}

{\small gabriel.lopes.cardoso@tecnico.ulisboa.pt, swapna.mahapatra@gmail.com, 
silvia.nagy@durham.ac.uk}
\\[0.6cm]
\end{center}

\begin{center} {\bf Abstract } 
\end{center}
\noindent
We consider the dimensional reduction to two dimensions of certain gravitational theories in $D \geq 4$ dimensions at the two-derivative level. It is known that the resulting field equations
describe an integrable system in two dimensions which can also be obtained by a dimensional reduction of the self-dual Yang-Mills equations in four dimensions. We use this relation to construct a single copy prescription for classes of gravitational solutions in  Weyl-Lewis-Papapetrou coordinates. 
In contrast with previous proposals, we find that the gauge group of the Yang-Mills single copy carries non-trivial information about the gravitational solution.
We illustrate our single copy prescription with various examples that include the extremal Reissner-Nordstrom solution, the Kaluza-Klein rotating attractor solution,  the Einstein-Rosen wave solution and the self-dual Kleinian Taub-NUT solution.

\vskip 5mm

\date{\today}

\section{Introduction}
In this paper, we consider certain gravitational theories in $D \geq 4$ dimensions at the two-derivative level, in the absence of a cosmological constant. We focus on solutions of the associated field equations that have 
sufficiently many commuting isometries so that they can also be regarded as solutions of the field equations that result from the dimensional reduction of these theories down to two dimensions using a two step procedure
\cite{Breitenlohner:1987dg,Nicolai:1991tt,Schwarz:1995af,Lu:2007jc}.
The resulting field equations in two dimensions are
partial differential equations (PDEs) in the so-called Weyl 
coordinates $(\rho,v)$, with $\rho >0$ and
$v \in \mathbb{R}$. As we will review in Section \ref{sec:dimred}, 
they are written in terms of a 1-form $A$ and take the form
\bea
d \left( \rho \star A \right) = 0 \;\;\;,\;\;\; F = d A + A \wedge A = 0\;.
\label{grfe2}
\eea
These field equations are known to
describe an integrable system in two dimensions \cite{Breitenlohner:1986um,Lu:2007jc}, namely, 
they are the 
compatibility condition for an auxiliary linear system
 of differential equations (a Lax pair), called the Breitenlohner-Maison (BM) linear system.

It is also known that the dimensional reduction of the self-duality equations for Yang-Mills fields in four dimensions can 
give rise to PDEs
in lower dimensions that describe integrable systems \cite{ward85,ablo}, and that a particular reduction to two dimensions gives the field equations \eqref{grfe2} \cite{ward85,Ivanova:1994du,mas}.

\begin{figure}[hbt!]
    \centering

\begin{tikzpicture}
 \node[block] at (6,2) (a) {PDEs in two dimensions: $d \left( \rho \star A \right) = 0 \;,\; F = dA + A \wedge A = 0 $}; 

\node[block] at (2,-1) (b) {Gravitational field equations in $D \geq 4$};

\node[block]  at (10,-1)  (c){Self-dual Yang-Mills equations in $D=4$};

\draw[line, blue, very thick] (b.north) -- (a.south);

\draw[line, blue, very thick] (c.north) -- (a.south);

\end{tikzpicture}

    \caption{Relating gravitational solutions in $D\geq 4$ to solutions of the self-dual Yang-Mills equations in $D=4$ through PDEs in two dimensions.}
        \label{fig:d24rel}
\end{figure}
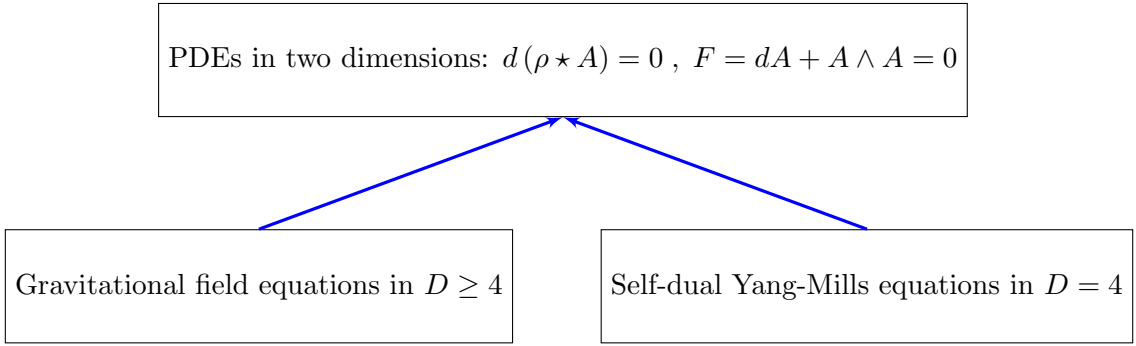

Thus, solutions to the PDEs \eqref{grfe2} give rise, on the one hand, to 
solutions of the gravitational field equations in $D \geq 4$, and on the other hand to solutions of the self-dual Yang-Mills equations in four dimensions. In this way, (non self-dual) solutions of the gravitational field equations in $D \geq 4$ can be related
to self-dual solutions of the Yang-Mills equations in four dimensions.
This is depicted in Figure \ref{fig:d24rel}.

Here, we will focus on two classes of solutions to the two-dimensional PDEs
\eqref{grfe2}, which we call class I and class II. These will be introduced in Section \ref{sec:dimred}.
Given a solution belonging to one of these classes, we give a procedure for constructing further solutions belonging to this class. 
Let us illustrate this by considering a class I solution. These are of the form 
\bea
A = d f \, C \, , 
\eea
where $f$ denotes a function
of the Weyl coordinates $(\rho,v)$ and $C$ denotes a constant matrix. By integrating $f$ over $v$, we obtain a new function $G(\rho,v)$, and therefore a 1-form 
\bea
A = dG \, C
\eea
that also solves \eqref{grfe2}, c.f. \eqref{fftG} and \eqref{solFT}. This procedure can, in principle, be repeated: integrating $G$ over $v$ will yield a new function from which one then constructs a 1-form $A$, as just described, that again solves \eqref{grfe2}.
Each of these solutions, obtained from $f$ by integration over $v$, will correspond to a solution 
of the gravitational theory in 
space-time dimensions $D \geq 4$.

We now wish to relate these gravitational solutions to solutions of the 
self-dual Yang-Mills equations in four dimensions in a physically meaningful way.
In the above discussion we have omitted the role played by the sources that source the gravitational solutions and 
the solutions to the self-dual Yang-Mills equations. 
We demand that the nature of the sources
be the same when mapping gravitational to non-gravitational solutions, that is, monopoles should be mapped to monopoles, and so on. This will serve as a guiding principle for constructing the single copy\footnote{The term single copy is taken from the literature on the classical double copy.}
of a gravitational solution that satisfies \eqref{grfe2}. To this end, in Section \ref{sec:scp}
  we focus on solutions to the PDEs \eqref{grfe2} that have both a class I and a class II description. We remark that one could set up the single copy prescription using only a class I
description, but additionally putting the solution in class II allows us to directly recover the scalar
version of the self-dual Yang-Mills solution.  
  Any such solution corresponds to a gravitational solution in space-time dimensions $D \geq 4$, and 
we describe how to construct a single copy description of these gravitational solutions. 
For illustrative purposes let us consider a gravitational solution whose class I description is of the form $A = df \, C$, as before. Given that we may construct new functions from $f$ by integrating $f$ over $v$, as explained above, which of the resulting 1-forms should we use to construct the single copy
of the original gravitational solution encoded in $A = df \, C$? This will depend on the nature of the sources of the gravitational solution. In most of the examples that we consider in this paper, the single copy is constructed using the function $G$ that is obtained by integrating $f$ once over $v$. This is summarized in Table \ref{fig:scp} for the case that the single copy gauge field $\A$ is a solution to the self-dual Yang-Mills equations in four-dimensional Minkowski space-time with line element
$ds^2 = - d U d V + d w d {\bar w}$, where the four-dimensional coordinates $(U, V, w , {\bar w })$ are expressed in terms of the Weyl coordinates $(\rho,v)$ by
$U = \tau - v, \, V = \tau + v, \, w = \rho \, e^{i \phi}, \, {\bar w} =  \rho \, e^{-i \phi} $.
Thus, the role of the associated solution $G$ is to serve as a bridge for obtaining the single copy description of the gravitational solution $A = df \, C$.

\begin{table}[h!]
\begin{center}
\begin{tabular}{|c | c | c | c|  } 
 \hline
  & $A$ &   & single copy ${\cal A}$ \\ [0.5ex] 
 \hline\hline 
 Class I & $d f \, C $ & $G(\rho, v) = \int^{v}_a f(\rho, {\tilde v}) \, d {\tilde v}  + g(\rho) $  & 
 \makecell{$\A_U = \A_w = 0$, \\  
 $\A_{\tau}= A_V = \frac14 f \, C \, , \, \A_{\bar w} = \frac12 \, e^{i \phi} \, \partial_{\rho} G \, C$ }\\ 
 [0.5ex] 
 \hline
 \end{tabular}
 \caption{Single copy description in a four-dimensional Minkowski space-time with line element $ds^2 = - d U d V + d w d {\bar w}$, where $U = \tau - v, \, V = \tau + v, \, w = \rho \, e^{i \phi}, \, {\bar w} =  \rho \, e^{-i \phi} $. }
 \label{fig:scp}
\end{center}
\end{table}

Among the explicit examples that we discuss, 
a notable exception to 
integrating $f$ only once over $v$ in order to obtain the single copy description of the gravitational solution is the Lorentzian version of the Eguchi-Hanson solution.
We find that in this example we have to integrate $f$ twice over $v$ in order to obtain the single copy
given in the literature \cite{Berman:2018hwd,Luna:2018dpt,Alawadhi:2019urr}. This is discussed in Section \ref{sec:egAnv}.

Let us summarise some of the salient features of our procedure.
Our approach is based on the field equations 
\eqref{grfe2}, which are given in Weyl coordinates. We assume that a solution of 
\eqref{grfe2} depends on a parameter $m$ which, when switched off, results in a 
gravitational solution that denotes a gravitational background.
We then work to first order in the parameter $m$ and consider the gravitational 1-form $A$ at this order. We further assume that at 
first order in $m$, $A$ has both a class I and a class II description, and we follow the steps outlined above for obtaining a single copy description of the gravitational solution.
We note that since our construction relates a gravitational solution that is (in general) not self-dual, to a self-dual solution of the Yang-Mills equations in four dimensions, half of the latter solution is not required when matching with results in the classical double copy literature\footnote{We note that a similar step appears in the so called Newman-Penrose map, which relates  self-dual solutions of Maxwell's equations to non-self-dual gravitational solutions, see \cite{Elor:2020nqe,Farnsworth:2023mff,Farnsworth:2021wvs}.}.
We confirm that the above procedure reproduces the single copy for various gravitational solutions discussed in the literature, such as the Schwarzschild solution \cite{Monteiro:2014cda}, the non-extremal Kerr solution \cite{Monteiro:2014cda}, the (Lorenzian version of the) self-dual Taub-NUT solution \cite{Luna:2015paa} and the (Lorenzian version of the) Eguchi-Hanson solution \cite{Berman:2018hwd,Luna:2018dpt,Alawadhi:2019urr}. We further illustrate our single copy procedure in the following examples: the extremal Reissner-Nordstrom solution, a solution describing a black hole in $AdS_2 \times S^2$, the Kaluza-Klein rotating attractor solution (which is supported by a non-constant dilaton field), the Einstein-Rosen wave solution and the self-dual Kleinian Taub-NUT solution, which is a four-dimensional gravitational solution with signature 
$(2,2)$ \cite{Crawley:2021auj}.

A noteworthy consequence of describing gravitational solutions in higher dimensions 
from a two-dimensional point of view through \eqref{grfe2}
is that gravitational solutions that look distinct in higher dimensions may turn out to be related to one another by solution generating transformations acting on $A$ (see for instance \cite{Camara:2017hez}). This 
in turn implies that their single copy descriptions are also related.  We will discuss two such examples, see Sections 
\ref{sec:exam12} and \ref{sec:scp}.

The paper is organized as follows.
In Section \ref{sec:dimred} we briefly review how the PDEs \eqref{grfe2} result from the two-step dimensional reduction of gravitational theories in $D\geq 4$. These PDEs are the compatibility conditions for a Lax pair, the Breitenlohner-Maison linear system. We give this Lax pair in terms of differential operators $({\cal L}, {\cal M})$ in two dimensions.  In Appendix \ref{sec:laxLM}
we give new 
 Lax pairs in 
three dimensions that also give rise to the field equations \eqref{grfe2} and which
provide an intermediate step between the previously known Lax pairs in four and two dimensions.
We then proceed and discuss two classes of solutions to the PDEs \eqref{grfe2}, which we denote by class I and class II, respectively. Given a solution in either one of these two classes, we construct further solutions in these classes by integration over the Weyl coordinate $v$. In Section \ref{sec:exam12} we discuss various known gravitational solutions that possess both a class I and a class II description.
In Section \ref{sec:EGHm} we discuss the class I and class II descriptions of the (Lorenzian version of the) Eguchi-Hanson solution. In Section \ref{sec:grYMm} we discuss the mapping of gravitational solutions belonging to either class I or II (or both) to solutions of the self-dual Yang-Mills equations in four dimensions. 
We do so by starting from the self-dual Yang-Mills equations and performing dimensional reductions of them to two dimensions.
In Appendix \ref{sec:2D4D} we present a different approach to this mapping, 
by starting on the gravitational side in two dimensions and expressing the 1-form $A$ in two dimensions as a 1-form in four space-time dimensions.
In Section \ref{sec:scp} we turn to the single copy description of gravitational solutions that have both a class I and a class II description, and we discuss the non-trivial role of the gauge group. We present our conclusions in Section \ref{sec:conc}. In Appendix 
\ref{sec:sdymds} we discuss the self-dual Yang-Mills equations in four dimensions in different space-time signatures, while in Appendix \ref{sec:asdymds} we give the anti self-dual Yang-Mills equations in signature $(1,3)$.

\section{The dimensionally reduced gravitational field equations \label{sec:dimred}}

We consider gravitational theories in $D \geq 4$ dimensions, at the two-derivative level, and without a cosmological constant term.
We focus on solutions to the associated field equations that only depend on two of the $D$ space-time coordinates. We assume that 
these gravitational theories, when dimensionally reduced to 
two dimensions using the two-step procedure 
discussed in \cite{Breitenlohner:1987dg,Nicolai:1991tt,Schwarz:1995af,Lu:2007jc}, 
take the following form as discussed below. Namely, 
in the first step, the theory is reduced to three dimensions, and subsequently, the 2-form fields are dualised into scalar fields. The resulting theory in three dimensions describes the coupling of a non-linear sigma model, whose target space is a symmetric space $G/H$,
to three-dimensional gravity. Then, a further reduction to two dimensions yields the following
non-linear field equations,
\bea 
\label{emotion2d}
d \left( \rho \star A \right) = 0 \;\;\;,\;\;\; \text{with} \; \; A = M^{-1} dM  \;,
\eea
where $M \in G/H$ is a coset representative of the symmetric space $G/H$. The latter
is invariant under an involution $\natural$ called generalized transposition, i.e.  $M^{\natural} = M $, and $M$
depends on two coordinates, denoted here by $\rho > 0$ and $v \in \mathbb{R}$, also called Weyl coordinates.
In \eqref{emotion2d}, 
$\star$ denotes the Hodge star operator in two dimensions,
\bea
\star d \rho = - \lambda \, dv \;\;\;,\;\;\; \star dv = d \rho \;\;\;,\;\;\; (\star)^2 = - \lambda \, {\rm id} \;,
\label{stho}
\eea
where $\lambda = \pm 1$ depends on the spacetime signature, as detailed below.

The matrix 1-form $A$ takes values in the Lie algebra of $G/H$. Denoting its generators by $\mathfrak{t}^a$, we have 
\bea
A = M^{-1} d M = A_a \mathfrak{t}^a \;\;\;,\;\;\; a= 1, \dots, n\;,
\label{decompA}
\eea
where the dependence on $(\rho,v)$ is encoded in the 1-forms $A_a= A_a (\rho,v)$.

Let us discuss the case 
when $D=4$. Then, a solution $M$ to \eqref{emotion2d} yields a gravitational solution, whose four-dimensional space-time metric in 
Weyl-Lewis-Papapetrou coordinates is a Weyl metric 
given by
\bea
ds_4^2 = - \lambda \, \Delta (dy  + B  d \phi  )^2 + \Delta^{-1} 
\left(  e^\psi \, ds_2^2 + \rho^2 d\phi^2 \right) \;.
\label{4dWLP}
\eea
The reduction from four down to two dimensions is performed along the directions $(y, \phi)$. If $\lambda =1$, the two-dimensional line element $ ds_2^2$ is space-like ($ds_2^2 = d\rho^2 + dv^2$), whereas if $\lambda =-1$, $ ds_2^2$ is time-like (either $ds_2^2 = - d\rho^2 + dv^2$ or 
$ds_2^2 = d\rho^2 - dv^2$). 
$\Delta, B$ are functions of $(\rho, v)$ determined  by the solution $M(\rho,v)$ of \eqref{emotion2d} and
$\psi(\rho,v)$ is a scalar function
determined  from  $M(\rho,v)$  by integration \cite{Lu:2007jc,Schwarz:1995af}, 
\bea
\partial_\rho \psi = \tfrac{1}{4} \, \rho \,  \text{Tr} \left(A_\rho^2 - \lambda \, A_v^2\right) \,,\qquad \partial_v \psi = \tfrac{1}{2} \, \rho \, 
 \text{Tr} \left(A_\rho A_v \right) \,. 
 \label{eq_psi}
\eea

As an example, consider the two-step reduction of the four-dimensional Einstein-Hilbert action to two dimensions.
The resulting coset is $G/H = SL(2, \mathbb{R})/SO(2)$, the involution $\natural$ is matrix transposition and the coset representative $M$ takes the form 
\bea
\label{M22DB}
M =
\begin{pmatrix}
 \Delta + \frac{{\tilde B}^2}{\Delta} &\;  \frac{{\tilde B}}{\Delta}\\
\frac{{\tilde B}}{\Delta}  &\;  \frac{1}{\Delta}
\end{pmatrix} \;,
\eea
where ${\tilde B}$ is related to $B$ through \cite{Breitenlohner:1986um}
\bea
 \rho \star d {\tilde B} = \Delta^2 \, dB \:.
 \label{tBB}
 \eea

Writing out the matrix 1-form $A$ as $A = A_{\rho} d\rho + A_v dv$, the field equations
\eqref{emotion2d} read
\bea
\partial_v A_v + \lambda  \partial_{\rho} A_{\rho} + \frac{\lambda}{\rho} A_{\rho} = 0,
\label{eoml1}
\eea
with 
\bea
F_{\rho v} = \partial_{\rho} A_v - \partial_v A_{\rho} + [A_{\rho}, A_v] = 0 
\label{FAA}
\eea
(i.e.  $F = d A + A \wedge A = 0$) 
by virtue of $A = M^{-1} dM$.

The field equations \eqref{eoml1} and \eqref{FAA} are the compatibility conditions for a Lax pair, the Breitenlohner-Maison linear system \cite{Breitenlohner:1986um,Lu:2007jc}. 
This Lax pair can be written in the form of a pair
of differential operators ${(\mathcal L}, {\mathcal M})$ 
acting on matrix functions $X$, 
\bea
{\mathcal L} X = 0 \;\;\;,\; \;\; {\mathcal M} X = 0 \;.
\label{LMX}
\eea
These differential operators are given by
 \bea
\mathcal{L} &=& - \partial_v + \tau \, D_{\rho} , \nonumber\\
 \mathcal{M} &=& \lambda \, \partial_{\rho} + \tau \, D_v \;\;\;,\;\;\; \lambda = \pm 1, 
 \label{LML}
 \eea
 where
 \bea
 D_. = \partial_. + A_. \;\;\;,\;\;\; A = M^{-1} d M,
 \eea
 and $\tau$ is a function of $\rho$ and $v$ as well as of a complex parameter $\omega$, which satisfies
 \bea
 \tau^2 - \lambda + \frac{2 \lambda \tau}{\rho} \left( \omega - v \right) = 0 \;\;\;,\;\;\; \omega \in \mathbb{C}. 
\label{relt2}
\eea
Likewise, the matrix functions $X$ in \eqref{LMX} are functions of 
$\rho, v, \omega$. The compatibility condition for the 
linear system \eqref{LMX},
\bea
[ {\mathcal L}, {\mathcal M}] X = 0 \;,
\eea
then implies the field equations \eqref{eoml1} and \eqref{FAA}.

The field equations \eqref{eoml1} and \eqref{FAA} can also be obtained as compatibility conditions for Lax pairs formulated in three and four dimensions.
The Lax pair in $D=4$ is determined by the self-dual Yang-Mills  equations in four dimensions, as we will discuss in Section \ref{sec:grYMm}.
In Appendix \ref{sec:laxLM} we give two Lax pairs in $D=3$ dimensions.
In $D=4$ dimensions the spectral parameter is constant, while in
$D=2$ dimensions the spectral parameter is non-constant. Interestingly, in $D=3$ dimensions one can describe the integrable system using either a constant or a non-constant spectral parameter.
We summarize our findings in Table \ref{table:1}.\\

\begin{table}[h!]
\begin{center}
\begin{tabular}{|c | c | c | c | c|} 
 \hline
 $D$ & spectral parameter & coordinates &  $L$  & $M$ \\ [0.5ex] 
 \hline\hline
 4 & $\Omega$ & $(U, V, w, {\bar w} )$ & $\partial_U   - \Omega {\mathcal D}_{\bar w} $ &  $\partial_w   + \Omega {\mathcal D}_V $ \\ 
 \hline
 3 & $\Omega$ & $ (\rho, v, \phi) $ & $  - e^{-2 i \phi} \, \partial_v + \frac12 \rho \, \Omega \,  \left(  D_{\rho} + \frac{i}{\rho} \, \partial_{\phi}  \right)  $ & $e^{- i \phi} \, \lambda \, \partial_{\rho} + \frac12 \rho \, \Omega \, e^{ i \phi} \,   D_{v}  $ \\
 \hline
 3 & $\tau (\rho, v, \omega)$ & $ (\rho, v, \phi) $ & $ e^{ i \phi} \left( -  \partial_v + \tau  \,  \left(  D_{\rho} + \frac{i}{\rho} \, \partial_{\phi}  \right) \right) $ & $ e^{- i \phi} \left(  \lambda \, \partial_{\rho} +  \tau \,   D_{v} \right)  $ \\
 \hline
 2 & $\tau (\rho, v, \omega)$  & $ (\rho, v) $ & $ - \partial_v + \tau \, D_{\rho} $ & $\lambda \, \partial_{\rho} + \tau \, D_v $ \\
 \hline
 \end{tabular}
 \caption{Lax pairs $(L,M)$ in various space-time dimensions $D$. $\Omega$ and $\omega$ denote constant spectral parameters, whereas $\tau$ is a non-constant spectral parameter that satisfies \eqref{relt2}.}
 \label{table:1}
\end{center}
\end{table}

A noteworthy consequence of describing gravitational solutions in higher dimensions through a coset representative $M$ is that gravitational solutions that look distinct in higher dimensions may turn out to be related to one another by solution generating transformations acting on $M$ (see \cite{Richterek:2004bb}, for instance). This is done by 
acting on $M$ with a suitably chosen constant matrix $g\in G/H$, to obtain the  matrix 
\bea
{\tilde M} = g^{\natural} M g \;, 
\eea
which results in a matrix 1-form 
\bea
{\tilde A} = g^{-1} A g \;.
\eea
An example thereof, discussed in \cite{Camara:2017hez}, is the solution generating transformation that maps the $3 \times 3$ matrix $M$ describing the near-horizon region of the extremal Reissner-Nordstrom black hole solution to the $3 \times 3$ matrix $\tilde M$ describing 
the full interpolating extremal Reissner-Nordstrom black hole solution in four space-time dimensions.
We refer to Section \ref{sec:exam12} for details on this solution generating technique. Then, as we will see in Section \ref{sec:scp}, the Harrison transformation is mapped to a non-abelian rotation of the Yang-Mills gauge field via our single copy procedure.

\subsection{Two classes of solutions}

In the following, we will consider two classes of solutions to \eqref{eoml1} and \eqref{FAA}.
Solutions belonging to the first class are of the gradient type, thus possessing a structure that is suggestive of the form of $A$, i.e. $A = M^{-1} d M$. Solutions in the second class take a form that results in a Plebanski type equation.

As we shall see, 
we define the two classes such that 
the role of \eqref{eoml1} and of \eqref{FAA} is interchanged between the two:
\eqref{FAA} is automatically satisfied by the solutions in class I, while \eqref{eoml1} is automatically satisfied by the solutions in class II.

\subsubsection{Class I}
The first class 
consists of solutions of the form 
\bea
A = (d f_i) C_i \;, 
\label{decompf}
\eea
where the $f_i$ are functions of $(\rho,v)$, and where the $C_i$ are constant commuting matrices that reside in a subsector of
the Lie algebra of $G/H$, so as to ensure that \eqref{decompf}
satisfies the field strength condition \eqref{FAA}.
Inserting this into \eqref{eoml1} and using that 
the constant matrices $C_i$ are linearly independent, gives
\bea
\partial^2_v f_i + \lambda  \partial^2_{\rho} f_i + \frac{\lambda}{\rho} \partial_{\rho} f_i = 0 \:.
\label{eomlfa}
\eea
When $\lambda = 1$, this is the harmonic equation in three-dimensional Euclidean space,
written in cylindrical coordinates $(\rho, v, \phi)$, acting on functions $f_i$ that are independent of $\phi$.
When $\lambda = -1$, this is the harmonic equation in three-dimensional Minkowski space.

\subsubsection{Class II}

The second class of solutions 
consists of solutions of the form 
\bea
A = - \frac{\lambda}{2}\, \star d P + \frac{P}{2 \rho} \, dv \;,
\eea
that is, 
\bea
A_{\rho} = - \frac{\lambda}{2} \partial_v P (\rho, v) \;\;\;,\;\;\; A_{v} = \frac12 \left( \partial_{\rho} + \frac{1}{\rho} \right) P (\rho, v) ,
\label{ArvP}
\eea
where 
\bea
P(\rho, v) = p_a (\rho,v) \, \mathfrak{t}^a .
\label{Pdecom}
\eea
This satisfies \eqref{eoml1}, while 
the condition \eqref{FAA}  imposes\footnote{We remark that the interaction term in \eqref{Pdiff} is of the form $[\{P,P\}_*]$, where $\{,\}_*$ is the modified Poisson bracket defined in \cite{SilvLip} in the context of self-dual gravity on an AdS background. Specifically, in \cite{SilvLip} we have a 4-dimensional theory in lightcone coordinates $(\mathfrak{u}, \mathfrak{v},\mathfrak{w},\bar{\mathfrak{w}})$, with 
\be 
\{f,g\}_*=\partial_{\mathfrak{w}} f \partial_\mathfrak{u} g- \partial_\mathfrak{u} f \partial_\mathfrak{w} g 
+\frac{2}{\mathfrak{u}-\mathfrak{v}}\left(f\partial_\mathfrak{w} g - \partial_\mathfrak{w} f  \, g \right) \;.
\ee 
If $f$ and $g$ are matrices, following \cite{SilvLip}, we define the combined Poisson-commutator bracket via
\be 
[\{f,g\}_*]=[\partial_\mathfrak{w} f ,\partial_\mathfrak{u} g]- [\partial_\mathfrak{u} f, \partial_\mathfrak{w} g] 
+[\frac{2}{\mathfrak{u}-\mathfrak{v}}f,\partial_\mathfrak{w} g] -[\partial_\mathfrak{w} f, \frac{2}{\mathfrak{u}-\mathfrak{v}}g] \;.
\ee
Note that because $\{,\}_*$ and $[,]$ are both anti-commuting, the combine bracket $[\{,\}_*]$ commutes, therefore we can take 
\be 
f=g=\frac{\sqrt{\lambda} P}{2}
\ee 
to get
\be 
[\{\frac{\sqrt{\lambda} P}{2},\frac{\sqrt{\lambda} P}{2}\}_*]=
-\frac{\lambda}{2}[\partial_\mathfrak{u} P,\partial_\mathfrak{w} P] +
\frac{\lambda}{2}[\frac{2}{\mathfrak{u}-\mathfrak{v}}P,\partial_\mathfrak{w} P] \;.
\ee 
The commutator terms in \eqref{Pdiff} will be obtained by 
setting
\bea
\mathfrak{u} = - \frac12 \rho + \frac12 \Sigma \;\;,\;\; \mathfrak{v} = - \frac52 \rho + \frac12 \Sigma 
\;\;,\;\; \mathfrak{w} = \frac12 v + \frac{i}{2} \hat{\Sigma} \;\;,\;\; \bar{\mathfrak{w}} = \frac12 v -\frac{i}{2} \hat{\Sigma} \;,
\eea
so that
\be 
\mathfrak{u}-\mathfrak{v}= 2 \rho,\quad \mathfrak{w}+\bar{\mathfrak{w}}=v \;,
\label{combfrak}
\ee
and using that $P$ only depends on the two combinations \eqref{combfrak}.
It would be interesting to explore the significance of this, particularly in light of recent deformations of the Breitenlohner-Maison system presented in \cite{LopesCardoso:2024tol}. We leave this for future work.}
\bea
\left( \partial^2_{\rho} + \lambda  \partial_v^2 + \frac{1}{\rho} \partial_{\rho}- \frac{1}{\rho^2}\right) P + \frac{\lambda}{2} [ \left( \partial_{\rho} 
+ \frac{1}{\rho} \right) P, \partial_v P ]= 0 ,
\label{Pdiff}
\eea
which can be written as
\bea
&& \left( \partial^2_{\rho} + \lambda  \partial_v^2 \right) P + \frac{\lambda}{2} [ \partial_{\rho} P, \partial_v P ] \nonumber\\
&& + \partial_{\rho} \left( \frac{P}{\rho} \right) +  \frac{\lambda}{2} [  \frac{P}{\rho} , \partial_v  P ] = 0 .
\label{PTE}
\eea

Let us focus on the case when the commutator terms in \eqref{PTE} vanish, in which case we obtain
\bea
\left( \partial^2_{\rho} + \lambda \partial_v^2 + 
\frac{1}{\rho} \partial_{\rho}  - \frac{1}{\rho^2} \right) P(\rho,v)= 0 \;.
\label{harmt}
\eea
Then, introducing 
\bea
\Psi (\rho, v, \phi) = \frac12 \, e^{i \phi} \, P(\rho,v) \;,
\label{PsiP}
\eea
we find that 
$\Psi (\rho, v, \phi)$ satisfies 
\bea
\left( \partial^2_{\rho} + \lambda \partial_v^2 + 
\frac{1}{\rho} \partial_{\rho}  + \frac{1}{\rho^2} \partial^2_{\phi}\right) \Psi (\rho,v,\phi)= 0 \:.
\label{difpsi}
\eea
When $\lambda = 1$, this is the harmonic equation in three-dimensional Euclidean space,
written in cylindrical coordinates $(\rho, v, \phi)$.
When $\lambda = -1$, this is the harmonic equation in three-dimensional Minkowski space.

\subsubsection{Solutions belonging to both classes}

There is a subspace of solutions to \eqref{eoml1} and \eqref{FAA} that allows for both a class I and a class II description. The examples that we will be discussing in the following sections all belong to this subspace of solutions.

Consider a solution of class I of the form
\bea
A = (d f) \, C \;,
\label{dfC}
\eea
with $C$ a constant matrix. Such a solution can also be written 
in the form 
\eqref{ArvP}, with $P$ given by
\bea
P (\rho,v) = p(\rho, v) \, C \;,
\label{Pp}
\eea
provided that $p$ satisfies \eqref{harmt}.
This is so, because the equations 
\bea
\partial_{\rho} f &=& - \frac{\lambda}{2} \, \partial_v p \,, \nonumber\\
\partial_v f &=& \frac12 \left( \partial_{\rho} + \frac{1}{\rho} \right) p 
\label{comclass12}
\eea
are compatible with one another provided $p$ satisfies \eqref{harmt}.
Similarly, given a solution of class II of the form \eqref{ArvP}, with $P$ given by
\eqref{Pp} and with $p$ satisfying \eqref{harmt}, it can be written in the form \eqref{dfC} by 
using \eqref{comclass12}.

A straightforward generalization of the above shows that solutions of class I of the form $A = (d f_i) C_i$, where the $C_i$ are commuting matrices, also admit a description as class II solutions. 

\subsection{Constructing further solutions \label{sec:fsol}}

Given a solution $f$ in class I or a solution $P$ in class II, we can construct further solutions
of \eqref{eomlfa} and of \eqref{harmt}, respectively, by integration with respect to $v$, as follows.

Let $f$ denote a function that satisfies \eqref{eomlfa}. Then we construct a function $G$ that also solves \eqref{eomlfa}, namely 
\bea
G(\rho, v) = \int^{v}_a f(\rho, {\tilde v}) \, d {\tilde v}  + g(\rho) \;,
\label{fftG}
\eea
where $g(\rho)$ has to satisfy
\bea
\lambda \, \rho \, g' (\rho) = - \int^{\rho}_b \, {\tilde \rho} \,\left( \partial_v f(v,{\tilde \rho})\right)_{\vert_{v = a}} \, d {\tilde \rho} \;.
\label{gp}
\eea
Here $a$ and $b$ denote constants.
$G - g$ is an indefinite integral in $v$, but also a parametric integral with parameter $\rho$. We assume that $f$ possesses the necessary continuity properties for applying the Leibniz integral rule for differentiation under the integral sign. 
Then, 
\bea
A = (d G) \, C \,,
\label{solFT}
\eea
with $C$ a constant matrix,
is a solution of \eqref{eoml1} and \eqref{FAA},
as can be easily verified by using \eqref{eomlfa} and \eqref{gp}.

Similarly, given a solution $\Psi$ of the form \eqref{PsiP}
that satisfies \eqref{difpsi},
we construct a function $H$ that also solves \eqref{difpsi},
namely 
\bea
H(\rho, v, \phi) = \frac12 e^{i \phi}  \left(\int^{v}_a P(\rho, {\tilde v}) \, d {\tilde v} +  h(\rho) \right) \;,
\label{HP}
\eea
where $h(\rho)$ has to satisfy
\bea
h'' + \frac{h'}{\rho} - \frac{h}{\rho^2} = - \lambda 
\left( \partial_v P \right)_{\vert v=a} \;.
\label{hdp}
\eea
$e^{-i \phi} \, H -  h$ is an indefinite integral in $v$, but also a parametric integral with parameter $\rho$. We assume that $P$ possesses the necessary continuity properties for applying the Leibniz integral rule for differentiation under the integral sign.
It can be easily verified that $H$ satisfies \eqref{difpsi}
by using \eqref{harmt} and \eqref{hdp}. Then,
\bea
A_{\rho} = - \lambda \, e^{-i \phi} \, \partial_v H = 
- \lambda \, e^{-i \phi} \, \Psi \;\;\;,\;\;\; A_{v} = 
e^{-i \phi} \left( \partial_{\rho} + \frac{1}{\rho} \right) H 
\label{ArvPsi}
\eea
is a solution of  \eqref{eoml1} and \eqref{FAA}.

Let us now consider a solution \eqref{dfC} that can also be written in the form 
\eqref{comclass12}.
We then introduce the functions $G$ and $H$ as above.
The associated solution \eqref{solFT} can then also be expressed in the form \eqref{ArvPsi}, since the relations 
\bea
e^{i \phi} \, \partial_{\rho} G \, C &=& - \lambda \partial_{v} H
=
- \frac{\lambda}{2} \, e^{i \phi} \, P \,, \nonumber\\
e^{i \phi} \, \partial_{v} G \, C= e^{i \phi} \,  f \, C &=& \left( \partial_{\rho} + \frac{1}{\rho} \right) H 
\label{relsGH}
\eea
are compatible by virtue of $H$ satisfying \eqref{harmt}.

An example thereof is provided by $\lambda =1$ with
\bea
f (\rho,v) = - 2 \frac{m}{\sqrt{\rho^2 + v^2}} \;\;\;,\;\;\; 
P (\rho,v) = - 4 m \left( \frac{v}{\rho \, \sqrt{\rho^2 + v^2}} \right) \, C \;.
\label{fp}
\eea
Using \eqref{fftG} and \eqref{HP}, we obtain
\bea
G(\rho,v) &=& m \, \log \left[ \frac{\sqrt{\rho^2 + v^2} - v}{\sqrt{\rho^2 + v^2} + v} \right] 
=-2m\ \tanh^{-1}\left(\frac{v}{\sqrt{\rho^2+v^2}} \right)\;,
\nonumber\\
H(\rho,v, \phi) &=& - 2 m \, e^{i \phi} \,  \frac{\sqrt{\rho^2 + v^2}}{\rho} \, C \;,
\label{GHfel}
\eea
which satisfy the differential equations \eqref{eomlfa} and \eqref{difpsi}, respectively.

\section{Examples \label{sec:exam12}}

We now give examples of gravitational solutions in four dimensions
that have both a class I and a class II description.

\subsection{Class I description}

 Let us consider solutions of the form \eqref{decompf} with one or two matrices $C_i$, 
\bea
A = df_i \, C_i \;,
\label{soldf}
\eea
where the $C_i$ are constant commuting matrices 
and the $f_i = f_i(\rho,v)$ satisfy \eqref{eomlfa}. 
Using \eqref{fftG},
to each $f_i$ we 
associate a 
function $G_i$ that satisfies the differential equation \eqref{eomlfa}. 

In all the examples discussed in this section, the solutions 
depend on various parameters which, when they are switched off, ensure
the vanishing of $A$.

There are two cases to consider, either $\lambda = 1$ or $\lambda =-1$ (c.f. \eqref{4dWLP}).
When $\lambda =1$, the examples that we consider are the Schwarzschild solution, 
the extremal Reissner-Nordstrom black hole in four space-time dimensions, 
a black hole in $AdS_2 \times S^2$,
the (Lorentzian version of the) self-dual Taub-NUT solution,  the non-extremal Kerr black hole, the rotating attractor solution for a Kaluza-Klein black hole. When $\lambda = -1$, one example that we consider is the Einstein-Rosen wave solution.  In addition, we also discuss the self-dual Kleinian Taub-NUT solution, which is a four-dimensional gravitational solution with signature $(2,2)$ \cite{Crawley:2021auj}.

\subsubsection{Exterior region of the Schwarzschild black hole
\label{subsec:Exterior region of the Schwarzschild black hole}}

The line element describing the exterior region of the Schwarzschild black hole solution can be
written in the form
\bea
ds_4^2 = - e^{- \varphi} dt^2 + e^{ \varphi} ds_3^2 
\label{schwarzsm}
\eea
with
\bea
ds_3^2 = dr^2 + (r^2 - m^2) \left( d \theta^2 + \sin^2 \theta \, d \phi^2 \right)
\eea
and
\bea
e^{- \varphi (r)} = \frac{ r -m}{r + m} \;\;\;,\;\;\; r > m \;.
\eea
This solution carries one parameter, the mass parameter $m$.

Introducing Weyl coordinates
\bea
\rho = \sqrt{r^2 - m^2} \, \sin \theta \;\;\;,\;\;\; v = r \cos \theta \;,
\eea
the line element  \eqref{schwarzsm} can be brought to the form 
\eqref{4dWLP} 
with (see, for instance, \cite{Aniceto:2019rhg}) 
\bea
\Delta (\rho, v) &=& \frac{v + m - \sqrt{(v +m )^2 + \rho^2}}{v - m - \sqrt{(v -m )^2 + \rho^2}} \;\;\;,\;\;\; B= 0 \;, \nonumber\\
\psi (\rho, v) &=& \log \left[ \frac12 \frac{v^2 + \rho^2 - m^2}{
\sqrt{(v +m )^2 + \rho^2} \sqrt{(v -m )^2 + \rho^2}} + \frac12\right] \;.
\eea
The associated matrix $M(\rho,v)$ 
(with $ M \in G/H = SL(2, \mathbb{R})/SO(2)$) 
is given by 
\eqref{M22DB},
with ${\tilde B} = 0$,
\bea
M (\rho, v) = 
 \begin{pmatrix}
 \Delta (\rho,v)  & 0 \\
 0 & \frac{1}{\Delta(\rho, v)}
 \end{pmatrix} \;\;\;,\;\;\; 
\Delta (\rho, v) = \frac{v + m - \sqrt{(v +m )^2 + \rho^2}}{v - m - \sqrt{(v -m )^2 + \rho^2}} \;,
\label{schmM}
\eea
which results in 
\bea
A = M^{-1} d M = df \, C
\eea
with
\bea
f (\rho, v) = \log \Delta (\rho,v) \;\;\;,\;\; C = \begin{pmatrix}
1 & 0 \\ 0 & -1
\end{pmatrix} \;.
\label{schwarzA}
\eea
To first order in the parameter $m$, $f$ reads
    \bea
    f(\rho,v) =  - 2 \frac{m}{\sqrt{\rho^2 + v^2}} \, ,
\label{fmontr}
\eea
which describes a monopole type potential. Since \eqref{fmontr}
is a solution of \eqref{eomlfa}
with $\lambda =1$, 
it will give rise to a solution \eqref{4dWLP}.
To it we associate the function
\bea 
G(\rho,v)=  m  \left[\log (\sqrt{\rho^2 + v^2} - v) - \log (\sqrt{\rho^2 + v^2} + v) \right] \;.
\eea

\subsubsection{Extremal Reissner-Nordstrom black hole
\label{sec:extRNh}}

The static extremal Reissner-Nordstrom black hole solution is a charged black hole solution of the Einstein+Maxwell theory in four space-time dimensions. It carries one electric charge $q$ and one magnetic charge $p$.
We introduce the combination $Q = q + i p$. In the near-horizon limit, the line element describes the 
product geometry $AdS_2 \times S^2$, and the solution is supported by  
electro-/magnetostatic potentials $\chi_e$ and $\chi_m$,
\bea
ds_4^2 &=&  - \frac{r^2}{|Q|^2} dt^2  + |Q|^2  \frac{dr^2}{r^2 }  + |Q|^2 \left( d \theta^2 + \sin^2 \theta \, d \phi^2 \right) \:, \nonumber\\
\chi_e + i \chi_m &=& \frac{Q}{|Q|^2} \, r \:.
\label{nearadshor}
\eea
Introducing Weyl coordinates
\bea
\rho = r \, \sin \theta \;\;\;,\;\;\; v = r \cos \theta \;,
\eea
the line element  \eqref{nearadshor} can be brought to the form 
\eqref{4dWLP} with $e^{\psi} = 1$, $B=0$ and $\Delta = (\rho^2 + v^2)/|Q|^2$.
This near-horizon solution is encoded in a $3 \times 3$ matrix 
 $M \in G/H = SU(2,1) / ( SL(2, \mathbb{R} ) \times U(1) ) $ given by \cite{Camara:2017hez}
\bea
 M = \begin{pmatrix}
 e^{\varphi} & \sqrt{2}  e^{\varphi} Z & 1 \\
 -\sqrt{2}  e^{\varphi} {\bar  Z} & -1 & 0 \\
 1 & 0 &  0
 \end{pmatrix} \;\;\;,\;\;\; e^{- \varphi} = \Delta \;\;\;,\;\;\; Z = \chi_e + i \chi_m = c \, e^{- \varphi/2} \;,
 \label{MMAdS}
 \eea
 where 
 \bea
 c = \frac{Q}{|Q|} \;\;\;,\;\;\; |c| = 1 \;.
 \eea
We note that the matrix $M$ can also be written as
\bea
 M = \frac{1}{|Z|^2} \begin{pmatrix}
 1 & \sqrt{2}  Z & |Z|^2 \\
 -\sqrt{2}   {\bar  Z} & -|Z|^2 & 0 \\
 |Z|^2 & 0 &  0
 \end{pmatrix} \;.
 \label{MZ}
 \eea
The associated matrix 1-form $A= M^{-1}dM$ reads $A = df \, C$ with 
\bea
f (\rho, v) = e^{\varphi (\rho,v)/2} = \frac{|Q|}{\sqrt{\rho^2 + v^2}}  \;\;\;,\;\; C 
= \begin{pmatrix} 
 0 & 0 & 0\\
 \sqrt{2} {\bar c} & 0 & 0 \\
 0 & \sqrt{2} c & 0
 \end{pmatrix} \;.
 \label{Ahatads}
 \eea
Note that $f$ is linear in the parameter $Q$ and that it describes a monopole type potential. The associated function $G$ is
\bea 
G(\rho,v)=  - \frac{|Q|}{2}  \left[\log (\sqrt{\rho^2 + v^2} - v) - \log (\sqrt{\rho^2 + v^2} + v) \right] \;.
\eea
The constant matrix $ C$ is a linear combination of two Lie algebra generators (given in (A.6) and (A.12) in \cite{Camara:2017hez}), 
\bea
{C} = \sqrt{2} \left( {\rm Re} \, c \right) \, (F_1 + F_2) - \sqrt{2} \left( {\rm Im} \, c \right) \, i(F_2 - F_1) \;.
\label{CEE}
\eea
When the black hole is electrically charged ($p=0$) we have $c =1$, while when it is magnetically charged ($q=0$) we have $c = i$.

The interpolating solution describing an extremal Reissner-Nordstrom black hole can be obtained from \eqref{MZ}
by applying a solution generating transformation called
Harrison transformation to $M$, as follows \cite{Camara:2017hez}. We act on $M$ with the transformation $g({\hat c})$, 
 \bea
 g({\hat c}) = \begin{pmatrix}
 1 & 0 & 0 \\
 - \sqrt{2} {\bar {\hat c}}  & 1 & 0 \\
 |{\hat c}|^2 &  - \sqrt{2} {\hat c} & 1
 \end{pmatrix} \;\;\;,\;\;\;  {\hat c} = - c/2 \;.
 \label{gc1}
 \eea
 The action on $M$ is given by ${\tilde M} = g^{\natural} ({\hat c}) M g({\hat c}) $. Here the involution ${\natural}$ is defined by 
 $ g^{\natural} ({\hat c}) = \eta g({\hat c})^{\dagger} \eta^{-1}$ with $\eta = {\rm diag} (1, -1, 1)$,
 \bea
 g^{\natural} ({\hat c}) = 
 \begin{pmatrix}
 1 & \sqrt{2} {\hat c} & |{\hat c}|^2 \\
 0 & 1 & \sqrt{2}  {\bar {\hat c}} \\
 0 & 0 & 1
 \end{pmatrix}.
 \label{MAdS}
 \eea
The resulting matrix $ {\tilde M} $ takes the form \eqref{MZ}, with $Z$ replaced by ${\tilde Z}$ \cite{Camara:2017hez},
\bea
Z \rightarrow {\tilde Z} = \frac{Z}{1- 2 {\bar {\hat c} Z}} = \frac{Z}{1 +  {\bar c} Z} \;.
\eea
Under \eqref{gc1}, the matrix 1-form $A$  transforms as
 \bea
 A =  M^{-1}d M \longrightarrow {\tilde A} = {\tilde M}^{-1}d {\tilde M} =  g^{-1} ({\hat c})  \, A \, g({\hat c}) ,
 \eea
and we obtain 
 \bea
 {\tilde A} = A \;.
 \eea
Thus, both the interpolating and the near-horizon solution have the same matrix 1-form $A$.

\subsubsection{Black hole in $AdS_2 \times S^2$
\label{Black hole in ADS2XS2}}

It is well known that the solution describing the near-horizon near-extremal limit of a Reissner-Nordstrom black hole in four space-time dimensions is a black hole in $AdS_2 \times S^2$ \cite{Sen:2008vm}.
Such a solution
can be obtained 
by applying a Harrison transformation (see Table 2 in \cite{Richterek:2004bb})   to the Schwarzschild solution, as discussed in \cite{Camara:2017hez}. Namely,
we embed the $2 \times 2$ matrix 
$M$ given in \eqref{schmM}
into a $3 \times 3$ matrix 
 $M \in G/H = SU(2,1) / ( SL(2, \mathbb{R} ) \times U(1) ) $,
 \bea
 M = 
 \begin{pmatrix}
 e^{- \varphi} & 0 \\
 0 & e^{\varphi}
  \end{pmatrix} \Longrightarrow
 M = \begin{pmatrix}
 e^{\varphi} & 0 & 0 \\
 0 & 1 & 0 \\
 0 & 0 &   e^{-\varphi}
 \end{pmatrix} ,
 \label{MMS}
 \eea
where $ e^{-\varphi} = \Delta$. 
The associated matrix one-form $A = M^{-1}d M$ reads
 \bea
 A = d f \, C \;\;\;,\;\;\; f = - \varphi \;\;\;,\;\;\; C = -\begin{pmatrix} 
 1 & 0 & 0\\
 0 & 0 & 0 \\
 0 & 0 & -1
 \end{pmatrix} \;.
 \label{A3schw}
 \eea
Now we act on $M$ with a transformation $g(c)$, as described below
\eqref{gc1}, i.e. ${\tilde M} = g^{\natural} ({c}) M g({ c})$ with 
$ g^{\natural} ({c}) = \eta g({c})^{\dagger} \eta^{-1}$ with $\eta = {\rm diag} (1, -1, 1)$ and
 \bea
 g(c) = \begin{pmatrix}
 1 & 0 & 0 \\
 - \sqrt{2} {\bar c}  & 1 & 0 \\
 |c|^2 &  - \sqrt{2} c & 1
 \end{pmatrix} \;\;\;,\;\;\; 
 c \in \mathbb{C} \;.
 \label{gc}
 \eea
We obtain
\bea
 {\tilde M} = \begin{pmatrix}
 \left( e^{\varphi/2} - |c|^2 \, e^{-\varphi/2} \right)^2 & \sqrt{2} \, c \left( 1 - |c|^2 \, e^{- \varphi} \right) &  |c|^2 \,e^{- \varphi}  \\
 -  \sqrt{2} \, {\bar c} \left( 1 - |c|^2 \, e^{- \varphi} \right) & 1 - 2 |c|^2 \, e^{- \varphi} & \sqrt{2} {\bar c} \, e^{- \varphi} \\
|c|^2 \,  e^{- \varphi} & - \sqrt{2} c \, e^{- \varphi} &  e^{- \varphi} 
 \end{pmatrix} = {\tilde M}^{\natural} .
 \eea
 Under \eqref{gc}, the 1-form $A$ in \eqref{A3schw}  transforms into
 \bea
 A =  M^{-1}d M \longrightarrow {\tilde A} = {\tilde M}^{-1}d {\tilde M} =  g^{-1} (c)  \, A \, g(c) .
 \eea
 Hence we obtain 
 \bea
 {\tilde A} = d {\tilde f}  \, {\tilde C} \;\;\;,\;\;\; {\tilde f}= f \;,
 \label{AfCt}
 \eea
 with ${\tilde C} =  g^{-1} (c)  \, C \, g(c)$,
where
\bea
 {\tilde C}  = - \begin{pmatrix} 
 1 & 0 & 0\\
 \sqrt{2} {\bar c} & 0 & 0 \\
 0 & \sqrt{2} c & -1
 \end{pmatrix} \;.
 \eea
Note that the dependence on the parameter $c$ is entirely contained in the matrix ${\tilde C}$, and that this dependence is linear in $c$. The matrix $\tilde C$ is a linear combination of Lie algebra generators,
\bea
{\tilde C} = - H_2 -  \sqrt{2} \left( {\rm Re} \, c \right)\, (F_1 + F_2) + \sqrt{2} \left( {\rm Im} \, c \right) \, i(F_2 - F_1) \;,
\eea
where $H_2, F_1 + F_2, i(F_2 -F_1)$ are the generators given in (A.6) and (A.12) in \cite{Camara:2017hez}.

If we now set $c= e^{i \alpha}$, 
  we get
 \bea
 {\tilde M} = \begin{pmatrix}
 \left( e^{\varphi/2} - e^{-\varphi/2} \right)^2 & \sqrt{2} \, e^{i \alpha} \left( 1 - e^{- \varphi} \right) &  e^{- \varphi}  \\
 -  \sqrt{2} \, e^{-i \alpha} \left( 1 - e^{- \varphi} \right) & 1 - 2 e^{- \varphi} & \sqrt{2} e^{- i \alpha} e^{- \varphi} \\
 e^{- \varphi} & - \sqrt{2} e^{ i \alpha} e^{- \varphi} &  e^{- \varphi} 
 \end{pmatrix} = {\tilde M}^{\natural} .
 \eea
The resulting solution has a line element of the form \eqref{schwarzsm}, with $e^{\varphi}$ replaced by
\bea
e^{\tilde \varphi} =  \left( e^{\varphi/2} -  e^{-\varphi/2} \right)^2 = \frac{4 m^2}{r^2 - m^2} \;.
\label{mettilp}
\eea
This line element describes a product geometry given by a black hole in $AdS_2 \times S^2$  (c.f.  (9.19) in \cite{Camara:2017hez}),
\bea
ds_4^2 = 4 m^2 \left( - (r^2 - m^2) dt^2  + \frac{dr^2}{r^2 - m^2} \right) + 4 m^2 \left( d \theta^2 + \sin^2 \theta \, d \phi^2 \right) \:,
\label{lineelads2s2}
\eea
where we have rescaled the time-like coordinate $t$ by a factor $4 m^2$.
This solution is supported by electric-/magnetostatic potentials $\chi_e / \chi_m$,
\bea
\chi_e + i \chi_m = \frac{c}{2m} \, (r -m ) \:.
\label{chiechim}
\eea

To lowest order in the parameter $m$, we infer using \eqref{AfCt}
that
\bea
{\tilde f} (\rho,v) = - 2 \frac{m}{\sqrt{\rho^2 + v^2}} \;.
\label{fmontr2}
\eea
The associated function $G$ is
\bea 
G(\rho,v) =  m  \left[\log (\sqrt{\rho^2 + v^2} - v) - \log (\sqrt{\rho^2 + v^2} + v) \right] \;.
\eea

\subsubsection{Self-dual Taub-NUT solution}
\label{sec:Self-dual Taub-NUT solution}

In Lorentzian signature, the Taub-NUT solution is given
by \cite{Taub:1950ez,Newman:1963yy} (see also (12.1) in \cite{Griffiths:2009dfa})
\bea
ds^2 = - f(r) \left( d {\bar t} - 2 \ell \cos \theta d \phi \right)^2 + \frac{dr^2}{f(r)} + (r^2 + \ell^2) \, \left( d\theta^2 + \sin^2 \theta \, d \phi^2 \right) 
\label{TNUT}
\eea
with 
\bea 
f(r ) = \frac{r^2 - 2 m r - \ell^2}{r^2 + \ell^2} .
\eea
Here, $m$ and $\ell$ are parameters, and $\ell$ is called NUT parameter.
Setting $\ell = + i m$, we obtain the self-dual Taub-NUT solution,
for which
\bea
f(r) = \frac{r-m}{r + m} \;.
\eea
We restrict to $r>m$ to ensure that $f(r) >0$.
Introducing Weyl coordinates (see (12.6) in \cite{Griffiths:2009dfa})
\bea
\rho = (r-m) \sin \theta \;\;\;,\;\;\; v = (r-m) \, \cos \theta, 
\label{weylco}
\eea
the line element for the self-dual Taub-NUT solution becomes
\bea
ds^2 = - \Delta (d {\bar t} + B \, d \phi)^2 + \Delta^{-1}
\left[ e^{\psi} \left( d \rho^2 + dv^2 \right) + \rho^2 \, d \phi^2 \right], 
\label{wlp}
\eea
where
\bea
\Delta =  \frac{\sqrt{\rho^2 + v^2}}{ \sqrt{\rho^2 + v^2 }+ 2 m } \;\;\;,\;\;\; e^{\psi} = 1 \;\;\;,\;\;\; B = - 2 i m \, \frac{v}{\sqrt{\rho^2 + v^2}} \;.
\label{DB}
\eea
Using \eqref{tBB} we infer
\bea
\partial_v {\tilde B} &=& \frac{\Delta^2}{\rho} \, \partial_{\rho} B = 2 i m \, \frac{v}{\sqrt{\rho^2  + v^2} \left( \sqrt{\rho^2  + v^2} + 2m \right)^2}, \nonumber\\
\partial_{\rho} {\tilde B} &=& - \frac{\Delta^2}{\rho} \, \partial_{v} B = 2 i m \, \frac{\rho}{\sqrt{\rho^2  + v^2} \left( \sqrt{\rho^2  + v^2} + 2m \right)^2},
\eea
which can be integrated and gives
\bea
{\tilde B} = i \left( \alpha - \frac{2  m }{\sqrt{\rho^2 + v^2} + 2m} \right) \;\;\;,\;\;\; \alpha \in \mathbb{C} .
\label{tB1}
\eea
The matrix $M(\rho,v)$  belonging to a complexification of $ M \in G/H = SL(2, \mathbb{R})/SO(2)$
 is obtained using \eqref{M22DB}, 
\bea
M (\rho,v) &=& \begin{pmatrix}
- (\alpha^2 -1) - \frac{2m (\alpha-1)^2}{\Phi}
& i \alpha + \frac{2 i (\alpha -1) m}{\Phi}\\
  i \alpha + \frac{2 i (\alpha -1) m}{\Phi}  & \frac{\Phi + 2m}{\Phi}
 \end{pmatrix} 
= \begin{pmatrix}
- (\alpha^2 -1) & i \alpha \\
i \alpha & 1 
\end{pmatrix}
+   \frac{2m}{\Phi} \, \hat{C} \;,\nonumber\\
\hat{C} &=&  \begin{pmatrix}
 - (\alpha -1)^2 &  i (\alpha -1) \\
 i (\alpha -1) & 1
 \end{pmatrix}
   \;\;\;,\;\;\; \Phi = \sqrt{ \rho^2 + v^2} \;.
   \label{MTN}
\eea
Note that $\det \hat{C} =0$. We obtain for $M^{-1}$,
\bea
M^{-1} (\rho,v) =  \begin{pmatrix}
1 & - i \alpha \\
- i \alpha & - (\alpha^2 -1) 
\end{pmatrix}
+   \frac{2m}{\Phi} \, {\tilde C} \;,\;
{\tilde C} &=&  \begin{pmatrix}
 1 & -  i (\alpha -1) \\
 - i (\alpha -1) & - (\alpha -1)^2 
 \end{pmatrix} \;,
   \label{MTN2}
\eea
with $\hat{C} {\tilde C} = 0$. This yields an exact class I description, 
\bea
A = M^{-1}d M = df \, C \; ,\;\ f(\rho,v) =  - \frac{2 m}{\sqrt{\rho^2 + v^2}}  \;,\; C = \begin{pmatrix}
1 - \alpha  & i \\
i (1-\alpha)^2 & \alpha -1
\end{pmatrix} \;,\; C^2 = 0 \;.
\eea
The expression for $f$ describes a monopole type potential.
The associated function $G$ is 
\bea 
G(\rho,v) =   m  \left[\log (\sqrt{\rho^2 + v^2} - v) - \log (\sqrt{\rho^2 + v^2} + v) \right] \;.
\eea

\subsubsection{The non-extremal Kerr black hole}

 The matrix $M(\rho,v)$ 
    (with $ M \in G/H = SL(2, \mathbb{R})/SO(2)$) that describes the 
exterior region of the non-extremal Kerr black hole with mass $m$ and angular momentum $a$  is obtained using \eqref{M22DB} 
and is given by \cite{Katsimpouri:2012ky,Camara:2017hez}
\bea
M(\rho, v)= \frac{1}{c^2 (u^2 - 1) + a^2 (y^2 - 1)}
\begin{pmatrix}
( c \, u - m)^2 +a^2 y^2  & 2 a m y \\
2 a m y & (c \, u+m)^2 +a^2 y^2
\end{pmatrix} \;,
\eea
where $c = \sqrt{m^2 - a^2}$, $u> 1, \, |y|<1$, with the prolate spheroidal coordinates $(u,y)$ expressed
in terms of the Weyl coordinates $(\rho, v)$ by (c.f. (G.6) in \cite{Harmark:2004rm})
\bea
u = \frac{ \sqrt{ \rho^2 + (v + c)^2} + \sqrt{ \rho^2 + (v - c)^2}}{2c} \;\;\;,\;\;\;
y =  \frac{ \sqrt{ \rho^2 + (v + c)^2} - \sqrt{ \rho^2 + (v - c)^2}}{2c} \;.
\eea
Computing $A = M^{-1} d M$ and expanding the result up to fifth order in the angular momentum parameter $a$ and to first
order in the mass parameter $m$ gives
\bea
A = d f_1 \, \begin{pmatrix}
1 & 0 \\
0 & -1
\end{pmatrix} + df_2 \begin{pmatrix}
0 & 1 \\
1 & 0
\end{pmatrix}
\label{kerrdf1df2}
\eea
with
\bea
f_1 (\rho,v) &=& m \left[ -  \frac{2}{\sqrt{\rho^2 + v^2}}  - a^2 \, \frac{\rho^2 - 2 v^2}{(\rho^2 + v^2)^{5/2}}
- a^4 \, \frac{3 \rho^4 - 24 \rho^2 v^2 + 8 v^4}{4 \, (\rho^2 + v^2)^{9/2}} \right] \;, \\
f_2 (\rho,v) &=& m \left[ a \, \frac{2  v}{(\rho^2 + v^2)^{3/2}}
+ a^3 \, \frac{ v (3 \rho^2 - 2 v^2)}{(\rho^2 + v^2)^{7/2}}
+ a^5 \, \frac{v (15 \rho^4 - 40 \rho^2 v^2 + 8 v^4)}{4 \, (\rho^2 + v^2)^{11/2}}
\right]
\;. \nonumber
\label{f1f2}
\eea
We note that $f_1$ carries even powers of the rotation parameter $a$, while $f_2$ carries odd powers of $a$.
When switching off $a$, $f_1$ reduces to the function $f$ of the Schwarzschild solution, to first order in $m$.

Since the two matrices in \eqref{kerrdf1df2} do not commute, we have to choose which one to keep in order to obtain a class I description. Since $f_2$ vanishes when switching off $a$, we keep the term proportional to $df_1$, which reduces to the function $f$ of the Schwarzschild solution when $a=0$.

We note that $f_1$ is the expansion of the function $- \phi$ given in eq. (52) of \cite{Monteiro:2014cda},
\bea
f_1 (\rho,v) = -2 m \, \frac{R^3(\rho,v)}{R^4(\rho,v) + a^2 v^2} \;\;\;,\;\;\; 
R = \frac{\sqrt{ \rho^2 + v^2 - a^2 + \sqrt{(\rho^2 + v^2 - a^2)^2 + 4 a^2 v^2}}}{\sqrt{2}} \;,
\label{f1R}
\eea
where $R$ is a solution of the equation (c.f. eq. (54) of \cite{Monteiro:2014cda})
\bea
\frac{\rho^2}{R^2 + a^2} + \frac{v^2}{R^2} = 1 \;,
\eea
and where $f_1$ 
satisfies the harmonic equation
\eqref{eomlfa} with $\lambda =1$.
The associated function $G$ is
\be 
G_1(\rho,v)=-2 m \tanh^{-1}\left(\frac{\sqrt{-a^2+\sqrt{a^4+2 a^2 \left(v^2-\rho ^2\right)+\left(\rho ^2+v^2\right)^2}+\rho ^2+v^2}}{\sqrt{2} v}\right) \;.
\ee 
When expanded to fourth order in the angular momentum parameter $a$, $G_1$ reads
\bea
G_1=-m\left [ \log (\sqrt{\rho^2 + v^2} + v) - \log (\sqrt{\rho^2 + v^2} - v)
+ \frac{a^2 v}{(\rho^2+v^2)^{3/2}} + \frac{a^4 v (3\rho^2 - 2 v^2)}
{4 (\rho^2+v^2)^{7/2}}
\right ] \;.
\eea

\subsubsection{ Kaluza-Klein black hole: rotating attractor}
\label{subsec: Kaluza-Klein black hole: rotating attractor}

Rotating Kaluza-Klein black hole solutions are solutions of the four-dimensional theory that is obtained by dimensional 
reduction of 
five-dimensional pure gravity on a circle. The resulting fields in four space-time dimensions are the metric, the
scalar field $\Phi$ and the Kaluza-Klein Maxwell field ${\cal A}_{\mu}$. We consider extremal rotating black hole solutions in this theory. The near-horizon behaviour of these solutions has been thoroughly studied in \cite{Astefanesei:2006dd} and it is determined by the attractor mechanism  in the presence of rotation (hence the name rotating attractors).
Let us focus on the near-horizon behaviour of slowly rotating extremal solutions (this is also referred to as the ergo-free branch). The near-horizon behaviour of the four-dimensional metric, the dilaton field $e^{- 4 \Phi/\sqrt{3}}$ and
the Maxwell field ${\cal F} = d {\cal A}$ is given in eqs. (5.36)-(5.44) of \cite{Astefanesei:2006dd}, and is determined in terms
of the electric-magnetic charges $(Q,P)$ (which, for definiteness, were taken to be positive in 
\cite{Astefanesei:2006dd})
and in terms of the angular momentum $J$. In particular, 
in the near-horizon region and in the presence of rotation, 
the dilaton field is not any longer constant,
\bea
e^{- 4 \Phi/\sqrt{3}} = \left( \frac{P}{Q} \right)^{2/3} \, \frac{ P Q - J \cos \theta}{ P Q +  J \cos \theta} \;.
\eea
The four-dimensional near-horizon metric can be written in the form \eqref{4dWLP} with $\lambda = 1$,
\bea
ds_4^2 = -  \Delta (dt  + B  d \phi  )^2 + \Delta^{-1} 
\left(  e^\psi \, ds_2^2 + \rho^2 d\phi^2 \right) \;,
\eea
with \cite{Camara:2017hez} 
\bea
\rho = r \sin \theta  \;\;,\;\; v = r \cos \theta \;\;\;,\;\;
\Delta = \frac{8 \pi (\rho^2 + v^2)}{\sqrt{P^2 Q^2 - J^2 \frac{v^2}{\rho^2 + v^2}}} \;\;,\;\;
 B = - \frac{J}{8 \pi} \, \frac{\rho^2}{(\rho^2 + v^2)^{3/2}} \;\;,\;\; e^{\psi } = 1 .
\eea
The matrix $M \in G/H$ that describes the near-horizon solution takes values in the coset 
$G/H = SL(3, \mathbb{R})/SO(2,1)$ and is given by \cite{Camara:2017hez}
\bea
M(\rho,v) = \frac{1}{\rho^2 + v^2} \begin{pmatrix}
- \frac{{\cal B}^2}{2{\cal D}} + \frac{2 {\cal A D} + {\cal B}^2}{2 {\cal D}} \frac{v}{\sqrt{\rho^2 + v^2}} \;\; &  {\cal B} \sqrt{\rho^2 + v^2} \;\; & {\cal C} (\rho^2 + v^2)  \\
-  {\cal B} \sqrt{\rho^2 + v^2} \;\; & {\cal D} (\rho^2 + v^2) \;\; & 0  \\
{\cal C} (\rho^2 + v^2) & 0 & 0 
\end{pmatrix} \;,
\eea
where
\bea
{\cal B}= \frac{1}{2 \sqrt{\pi} } \, P^{1/3} \, Q^{2/3} \, , \,  {\cal C}= - \left( \frac{P}{Q} \right)^{1/3} \, , \,
{\cal D} = - \left( \frac{Q}{P} \right)^{2/3} \, , \,  2 {\cal A D} + {\cal B}^2 = \frac{1}{4 \pi } \, \left( \frac{Q}{P} \right)^{1/3} 
  J \, , \nonumber\\
\eea
with ${\cal B}/{\cal D} \propto P, \, {\cal B}/{\cal C} \propto Q$ and 
$(2 {\cal A D} + {\cal B}^2)/({\cal C D}) \propto J$.
These constants satisfy $- {\cal C}^2 {\cal D} = 1$.

The associated matrix $A = M^{-1} d M$ reads
\bea
A = df_1 \, C_1 + df_2 \, C_2 \;,
\label{f12ra}
\eea
where $C_1$ and $C_2$ denote commuting matrices given by
\bea
C_1 = \begin{pmatrix}
0 & 0 & 0 \\
- \frac{{\cal B}}{{\cal D}} & 0 & 0 \\
0 & \frac{{\cal B}}{{\cal C}} & 0 
\end{pmatrix} \;\;\;,\;\;\; C_2 = \begin{pmatrix}
0 & 0 & 0 \\
0 & 0 & 0 \\
- \frac{ 2 {\cal A D} + {\cal B}^2 }{2 {\cal C D} } & 0 & 0
\end{pmatrix} \;,
\label{matC1C2}
\eea
and where the functions $f_1$ and $f_2$ are given by
\bea\label{KKf1f2}
f_1 (\rho, v) = \frac{1}{\sqrt{\rho^2 + v^2}} \;\;\;,\;\;\; f_2 (\rho, v) = 
- \frac{v}{(\rho^2 + v^2)^{3/2} } \;.
\eea
The associated functions $G_1$ and $G_2$ take the form
\bea\label{KKG1G2}
G_1 (\rho,v)= \frac12 \left[\log (\sqrt{\rho^2 + v^2} + v) - \log (\sqrt{\rho^2 + v^2} - v) \right] \;\;\;,\;\;\;
G_2 (\rho,v) = \frac{1 }{{\sqrt{\rho^2+v^2}}} \;.
\eea

Note that \eqref{f12ra} is an example of the form \eqref{decompf} based on two commuting matrices.
The matrices $C_1$ and $C_2$ are linear in $Q, P, J$.
When $J=0$, the matrix $C_2$ vanishes, and \eqref{f12ra} reduces to the case of a static attractor solution based on $AdS_2 \times S^2$.

\subsubsection{Einstein-Rosen wave}

The Einstein-Rosen wave describes a cylindrical gravitational wave.
In Weyl coordinates, its line element takes the form \eqref{4dWLP} with
$\lambda = -1$
(see, for instance, \cite{Penna:2021kua,Camara:2022gvc}) and 
\bea
\Delta (\rho, v) &=& e^{f(\rho, v)} \;\;\;,\;\;\; 
f(\rho,v) = 2 \cos (k v) \, J_0(k \rho) - 2 \;\;\;,\;\;\; 
B = 0 \;, \nonumber\\
\psi (\rho, v) &=& k^2 \rho^2 J^2_0(k \rho) + k^2 \rho^2 J^2_1(k \rho)
- 2 k \cos^2 (k v) \, \rho \, J_0(k \rho) J_1(k \rho) \:,
\eea
where $J_0$ and $J_1$ denote Bessel functions of the first kind. The parameter $k$ takes values in $\mathbb{R}_0^+$. When $k=0$ we recover flat space-time, since $J_0(0) = 1$.

The matrix $M(\rho,v)$ (with $ M \in G/H = SL(2, \mathbb{R})/SO(2)$) 
is given by \eqref{M22DB} with ${\tilde B} = 0$, 
\bea
M (\rho, v) = 
 \begin{pmatrix}
 \Delta (\rho,v)  & 0 \\
 0 & \frac{1}{\Delta(\rho, v)}
 \end{pmatrix} \;\;\;,\;\;\; 
\label{ersol}
\eea
and results in $A = d f \, C$ with 
\bea
C = \begin{pmatrix}
1 & 0 \\ 0 & -1
\end{pmatrix} \;.
\eea
The associated function $G$ is
\bea
G(\rho,v) = 
\frac{2}{k} \sin (k v ) \, J_0(k \rho) - 2v \;.
\eea
To lowest order in $k$ we obtain 
\bea
f(\rho,v) &=& k^2 \left( -  v^2 J_0(0) +  \rho^2 J''_0(0) \right)
=
k^2 \left( - v^2  - \frac12  \rho^2 \right) \;, \nonumber\\
G(\rho,v) &=& 
k^2 \left( - \frac13 v^3  - \frac12 v \rho^2 \right) \;,
\label{fGk2}
\eea
where we used $J_0(0) = 1, J''_0(0) = - \frac12$.

\subsubsection{Self-dual Kleinian Taub-NUT solution}
\label{sec:Self-dual Kleinian Taub-NUT solution}

All the four-dimensional gravitational solutions described above have signature $(1,3)$ and they have a line element of the form given in \eqref{4dWLP}. Here we consider a four-dimensional gravitational solution with signature $(2,2)$, namely the self-dual Kleinian Taub-NUT solution discussed in \cite{Crawley:2021auj}, whose line element takes the form 
\bea
ds^2 =  f(r) \left( d {\bar t} - 2 m \cosh \theta d \phi \right)^2 + \frac{dr^2}{f(r)} - (r^2 - m^2) \, \left( d \theta^2 + \sinh^2 d\phi^2 \right) \;,
\label{TNUTK}
\eea
where we have set $\ell = + i m$ and 
\bea
f(r) = \frac{r-m}{r + m} \;.
\eea
We restrict to $r>m$ to ensure that $f(r) >0$.
Introducing Weyl coordinates 
\bea
\rho = (r-m) \sinh \theta \;\;\;,\;\;\; v = (r-m) \, \cosh \theta, 
\label{weylcok}
\eea
the line element for the self-dual Kleinian Taub-NUT solution 
takes a form that differs from \eqref{4dWLP} by signs, namely
\bea
ds^2 =  \Delta (d {\bar t} + B \, d \phi)^2 + \Delta^{-1}
\left[ e^{\psi} \left( dv^2 - d \rho^2  \right) - \rho^2 \, d \phi^2 \right], 
\label{wlpk}
\eea
where
\bea
\Delta =  \frac{\sqrt{v^2 - \rho^2}}{ \sqrt{v^2 - \rho^2 }+ 2 m } \;\;\;,\;\;\; e^{\psi} = 1 \;\;\;,\;\;\; B = - 2  m \, \frac{v}{\sqrt{v^2 - \rho^2 }} \;.
\eea
Note that $v^2 - \rho^2 >0$ in view of $r>m$.
Using \eqref{tBB} with $\lambda =-1$, 
\bea
\partial_v {\tilde B} &=& \frac{\Delta^2}{\rho} \, \partial_{\rho} B 
, \nonumber\\
\partial_{\rho} {\tilde B} &=&  \frac{\Delta^2}{\rho} \, \partial_{v} B ,
\eea
we obtain
\bea
{\tilde B} =   \alpha + \frac{2  m }{\sqrt{v^2 - \rho^2 } + 2m}  \;\;\;,\;\;\; \alpha \in \mathbb{C} .
\eea
We define the  matrix $M(\rho,v)$ by 
\bea
\label{M22DBK}
M =
\begin{pmatrix}
 \Delta - \frac{{\tilde B}^2}{\Delta} &\; i \frac{{\tilde B}}{ \Delta} \\
i \frac{{\tilde B} }{\Delta}  &\;  \frac{1}{\Delta}
\end{pmatrix} \;,
\eea
which satisfies $M = M^T$ as well as
\bea
\det M = 1 \;.
\eea
Then we obtain,
\bea
M (\rho,v) &=& \begin{pmatrix}
1 - \alpha^2 & i \alpha \\
i \alpha & 1 
\end{pmatrix}
+   \frac{2m}{\tilde \Phi} \, \hat{C} \;,\nonumber\\
\hat{C} &=&  \begin{pmatrix}
 - (1 + \alpha )^2 &  i (1+\alpha ) \\
 i (1+\alpha ) & 1
 \end{pmatrix}
   \;\;\;,\;\;\; {\tilde \Phi} = \sqrt{ v^2 - \rho^2 } \;.
 \eea
Note that $\det \hat{C} =0$. We obtain for $M^{-1}$,
\bea
M^{-1} (\rho,v) =  \begin{pmatrix}
1 & - i \alpha \\
- i \alpha & 1 - \alpha^2  
\end{pmatrix}
+   \frac{2m}{\tilde \Phi} \, {\tilde C} \;,\;
{\tilde C} &=&  \begin{pmatrix}
 1 &  - i (1 + \alpha) \\
 - i (1 + \alpha ) & - (1+\alpha )^2 
 \end{pmatrix} \;,
 \eea
with $\det  {\tilde C} = 0$, $\hat{C} {\tilde C} = 0$. It follows that
\bea
A = M^{-1}d M = df \, C \; ,\;\ f(\rho,v) = - \frac{2 m}{\sqrt{v^2 - \rho^2}}  \;,\; C = \begin{pmatrix}
1+\alpha & -i \\
-i (1+\alpha)^2 & -(1+\alpha)
\end{pmatrix} \;,\; C^2 = 0 \:.
\eea
This expression is exact. The
associated function $G$ is 
\bea 
G(\rho,v) =  - m  \left[ \log (\sqrt{v^2 - \rho^2} + v) - 
\log (\sqrt{v^2 - \rho^2 } - v)  \right] \;.
\eea
It satisfies \eqref{eomlfa} with $\lambda = -1$.

\subsection{Class II description}

 Let us now turn to the class II description of the solutions given above.

We bring these solutions into the form \eqref{ArvP} and \eqref{Pdecom} with one or two matrices $C_i$.
Using \eqref{HP},
to each $P_i$ we 
associate a 
function $H_i$ that satisfies the differential equation \eqref{difpsi}.

\subsubsection{Exterior region of the Schwarzschild black hole}

Using \eqref{ArvP}, 
we associate the following matrix $P$ to $A$,
\bea
P(\rho, v) = \frac{2}{ \rho} \left(  \sqrt{(v - m )^2 + \rho^2}- \sqrt{(v + m )^2 + \rho^2} \right) \, C \;\;\;,\;\;\;
C = 
\begin{pmatrix}
1 & 0 \\ 0 & -1
\end{pmatrix} .
\label{schwarzs}
\eea
To first order in $m$ we obtain
\bea
P (\rho, v) = - 4 m  \,  \frac{v}{\rho \,  \sqrt{\rho^2 + v^2}} \, C \;,
\label{Pschwarzm}
\eea
and the associated function $H$ reads
\bea
H (\rho, v, \phi) = - 2 m \, e^{i \phi} \, \frac{\sqrt{\rho^2 + v^2}}{ \rho} \, C \;.
\eea

\subsubsection{Extremal Reissner-Nordstrom black hole}

We associate a matrix $P$ to ${\tilde A}$ using 
 \eqref{ArvP}, 
 \bea
 {\tilde A}_{\rho} &=&  g^{-1} (c)  \, A_{\rho}  \, g(c)   = A_{\rho} =  - \frac12 \partial_v P
  (\rho, v) \;, \nonumber\\
  {\tilde A}_{v} &=&  g^{-1} (c)  \, A_{v}  \, g(c) = A_v =  \frac12 \left( \partial_{\rho} + \frac{1}{\rho} \right) P
  (\rho, v) \; , 
\eea
which results in
  \bea
     P (\rho,v) = 2 \, |Q|  \frac{v}{\rho \sqrt{v^2 + \rho^2}} \, C
     \;\;\;,\;\;\;  C 
= \begin{pmatrix} 
 0 & 0 & 0\\
 \sqrt{2} {\bar c} & 0 & 0 \\
 0 & \sqrt{2} c & 0
 \end{pmatrix} \;\;\;,\,\;\;\; c = \frac{Q}{|Q|} \;.
      \eea
The associated function $H$ reads
\bea
H (\rho, v, \phi) =  |Q|  \, e^{i \phi} \, \frac{\sqrt{\rho^2 + v^2}}{ \rho} \, C \;.
\eea

\subsubsection{Black hole in $AdS_2 \times S^2$}

The associated matrix $P$ is
\bea
P(\rho, v) =
\frac{2}{ \rho} \left(  \sqrt{(v - m )^2 + \rho^2}- \sqrt{(v + m )^2 + \rho^2} \right) \, C \;\;,\;\;
 C  = - \begin{pmatrix} 
 1 & 0 & 0\\
 \sqrt{2} e^{- i \alpha} & 0 & 0 \\
 0 & \sqrt{2} e^{i \alpha} & -1
 \end{pmatrix} \;,
 \eea
 where $\alpha \in \mathbb{R}$.
To first order in $m$ we obtain
\bea
P (\rho, v) = - 4 m  \,  \frac{v}{\rho \,  \sqrt{\rho^2 + v^2}} \, C \;,
\eea
as well as
\bea
H (\rho, v, \phi) = - 2 m \, e^{i \phi} \, \frac{\sqrt{\rho^2 + v^2}}{ \rho} \, C \;.
\eea

\subsubsection{Self-dual Taub-NUT solution}

The associated matrix $P$ is
\bea
P (\rho, v) = - 4 m  \,  \frac{v}{\rho \,  \sqrt{\rho^2 + v^2}} \, C 
\;\;\;,\;\;\; C = \begin{pmatrix}
1 - \alpha  & i \\
i (1-\alpha)^2 & \alpha -1
\end{pmatrix} \;\;\;,\;\;\; C^2 = 0 \;,
\eea
where $\alpha \in \mathbb{C}$, 
and 
\bea
H (\rho, v, \phi) = - 2 m \, e^{i \phi} \, \frac{\sqrt{\rho^2 + v^2}}{ \rho} \, C \;.
\eea

\subsubsection{The non-extremal Kerr black hole}

To first order in $m$ and to fourth order in the rotation parameter $a$, the function $P$ associated to $f_1$ given in \eqref{f1f2}
reads
\bea
P_1 (\rho, v) =-m\left [\frac{4v}{\rho \sqrt{(\rho^2+v^2)}} +\frac{6 a^2 \rho v}
{(\rho^2+v^2)^{5/2}} + \frac{5a^4 \rho v (3\rho^2-4v^2)}{2(\rho^2+v^2)^{9/2}}
\right ] \,C \;\;,\;\; C =  \begin{pmatrix}
1 & 0 \\
0 & -1
\end{pmatrix} \; .
\eea
The associated function $H_1$ is
\bea
H_1 (\rho, v, \phi) = -  m \, e^{i \phi} \left( 2  \frac{\sqrt{\rho^2 + v^2}}{ \rho} - 
  a^2 \frac{\rho}{(\rho^2 + v^2)^{3/2}}
-  \frac{a^4}{4} \rho \frac{( \rho^2 - 4 v^2 ) }{(\rho^2 + v^2)^{7/2}}
\right)  C \;.
\eea

\subsubsection{Kaluza-Klein black hole: rotating attractor}

The associated matrix $P$ (c.f. \eqref{Pdecom}) takes the form
\bea
P (\rho,v)  = p_1(\rho,v) \, C_1 + p_2(\rho,v) \, C_2
\eea
with
\bea
p_1(\rho,v) = 2 \frac{v}{\rho \sqrt{\rho^2 + v^2}} \;\;\;,\;\;\; 
p_2 (\rho,v) = \frac{2 \rho}{(\rho^2 + v^2)^{3/2}} \;.
\eea
The matrices $C_1$ and $C_2$ are given in \eqref{matC1C2}.
The associated functions $H_1$ and $H_2$ are
\bea
H_1 (\rho, v, \phi)= e^{i \phi} \,\frac{ {\sqrt{\rho^2+v^2}}}{\rho} \, C_1 \;\;\;,\;\;\;
H_2 (\rho, v, \phi)= e^{i \phi} \frac{v}{\rho \sqrt{\rho^2+v^2}} \, C_2 \;.
\eea

\subsubsection{Einstein-Rosen wave}

The associated matrix $P$ is
\bea
P(\rho, v) = \frac{4}{k}  \, \sin  ( k  v ) 
\,  J'_0( k  \rho) \, C \;\;\;,\;\;\; C = 
\begin{pmatrix}
1 & 0 \\ 0 & -1
\end{pmatrix} 
\label{einro}
    \eea
and 
\bea
H (\rho, v, \phi) = e^{i \phi} \left(  - \frac{2}{k^2} 
 \cos (k v) 
\,   J'_0 (k \rho) -  \rho \right) \, C \;.
\eea
To lowest order in $k$ we obtain 
\bea
H (\rho, v, \phi) = - k^2 \, e^{i \phi} \left( \frac12 v^2 \rho + \frac18 \rho^3 \right) \;.
\eea
$P$ satisfies 
\bea
\left( \partial^2_{\rho} - \partial_v^2 + \frac{1}{\rho} \partial_{\rho} - \frac{1}{\rho^2} \right) P(\rho, v) = 0
\label{harmB}
\eea
by virtue of the differential equation for the Bessel function $J_0 (\rho)$,
\bea
J_0''(\rho)  + \frac{J_0' (\rho)}{\rho} + J_0 (\rho)= 0 \;.
\eea

\subsubsection{Kleinian self-dual Taub-NUT solution}

The associated matrix $P$ is
\bea
P (\rho, v) =  4 m  \,  \frac{v}{\rho \,  \sqrt{v^2 - \rho^2 }} \, C 
\;\;\;,\;\;\; C = \begin{pmatrix}
1 + \alpha & -i \\
-i (1+\alpha)^2 & -(1+\alpha)
\end{pmatrix} \;\;\;,\;\;\; C^2 = 0 \;,
\eea
where $\alpha \in \mathbb{C}$, and 
\bea
H (\rho, v, \phi) =   2 m \, e^{i \phi} \, \frac{\sqrt{v^2 - \rho^2 }}{ \rho} \, C \;.
\eea
The latter satisfies \eqref{difpsi} with $\lambda = -1$.

\section{Another example: the Eguchi-Hanson metric \label{sec:EGHm}}

In this section we discuss the Eguchi-Hanson metric in Weyl coordinates. Subsequently we determine the associated matrix 1-form $A = M^{-1} d M$. 

The Eguchi-Hanson metric depends on one parameter, $a$.
When switching off $a$, the line element describes four-dimensional flat space-time. 
We have verified that there does not exist a coordinate change for the Eguchi-Hanson metric which takes us to Weyl coordinates in such a way that (i) the background metric is mapped to the vacuum version of the Weyl metric with $A_\rho=A_v=0$, and (ii) the linear fluctuation (i.e. $\mathcal{O}(a^4)$) is in the standard Weyl form. In order to proceed, we will relax condition (i), while still requiring that (ii) is satisfied, and describe the corresponding coordinate change below. We will then provide a single copy prescription for the linear part, as follows.
We expand $A$ in powers of the parameter $a^4$, i.e. $A = A^{(0)} + A^{(1)} + {\cal O} (a^8)$, where 
$A^{(0)}$ (which is non-vanishing) describes the background.
We focus on $A^{(1)}$ and show that it admits both a class I and a class II description,
based on functions $f_1$ and $p_1$, respectively.
In Section \ref{sec:scp} we will show that the single copy description for the Eguchi-Hanson solution is constructed
using $p_1$ as a starting point and integrating it \underline{twice} with respect to the coordinate $v$. 
We denote the resulting function by ${\hat H}_1$. This differs from the examples discussed in the previous section, 
since for these the single copy description is based on a function $H$ that is obtained from the function $p$ by integrating
\underline{once} with respect to $v$.  The number of integrations over $v$ that one has to perform to obtain a single copy description differs from solution to solution and is related to the nature of the sources supporting the gravitational solution.

The Eguchi-Hanson metric is a self-dual Euclidean solution described by \cite{Eguchi:1978xp}
\bea
ds^2_4 = \frac14 \, r^2 \, h(r) \left( d \gamma + \cos \theta \, d \phi \right)^2 + \frac{dr^2}{h(r)} + \frac14 \, r^2 \, \left( d \theta^2 + \sin^2 \theta \, d \phi^2 \right) \;\;\;,\;\;\; 0 \leq \gamma < 4 \pi, 
\label{ehm}
\eea
with
\bea
h(r) = 1 - \frac{a^4}{r^4} > 0.
\eea

We now write the Eguchi-Hanson metric in Weyl form 
\bea
ds^2 =  \Delta (d \gamma + B \, d \phi)^2 + \Delta^{-1}[ e^{\psi} \left( d \rho^2 + dv^2 \right) + \rho^2 \, d \phi^2 ]  .
\label{euclwlp}
\eea
To do so, we first 
compare \eqref{euclwlp} with \eqref{ehm} and infer
\bea
\Delta &=& \frac14 \, r^2 \, h(r) \, \nonumber\\
B &=& \cos \theta  \nonumber\\
\rho^2 &=&  \frac14 \, r^2 \, \Delta \, \sin^2 \theta .
\eea
Hence,
\bea
\rho = \frac14 \, r^2 \, \sqrt{h} \, \sin \theta .
\eea
For convenience, we define 
\bea
F(r) =  \frac14 \, r^2 \, \sqrt{h(r)},
\label{Ff}
\eea
so that
\bea
\rho = F(r) \, \sin \theta.
\eea
We introduce
\bea
v = G(r) \, \cos \theta,
\eea
and calculate
\bea
d\rho^2 + d v^2 &=& dr^2 \left( (F')^2 \, \sin^2 \theta + (G')^2 \, \cos^2 \theta \right) + d \theta^2 \left( F^2 \, \cos^2 \theta + G^2 \, \sin^2 \theta \right) \nonumber\\
&& + 2 dr d \theta \left( F' \, F - G' \, G \right) \sin \theta \, \cos \theta.
\label{rvFG}
\eea
Since there are no terms proportional to $ \sin \theta \, \cos \theta$ in \eqref{ehm}, we demand
\bea
F' \, F - G' \, G = 0 \longleftrightarrow  ( F^2)' = (G^2)' \longrightarrow G^2 = F^2 + b \;\;\;,\;\;\; b \in \mathbb{R} \;.
\eea
Hence
\bea
G(r) = \sqrt{F^2(r) + b } .
\label{G}
\eea
Comparing \eqref{rvFG} with \eqref{ehm} results in
\bea
&& dr^2:  \;\;\; \left( (F')^2 \, \sin^2 \theta + (G')^2 \, \cos^2 \theta \right)  \frac{e^{\psi}}{\Delta}  = \frac{1}{h}, \nonumber\\
&& d \theta^2: \;\;\; \left( F^2 \, \cos^2 \theta + G^2 \, \sin^2 \theta \right) \frac{e^{\psi}}{\Delta} = \frac{r^2}{4} .
\label{drdtcond}
\eea
By combining these two equations we infer
\bea
\left( \frac{4}{r^2} \, F^2 - h ( G')^2 \right) \cos^2 \theta = \left( h (F')^2 -  \frac{4}{r^2} \, G^2 \right) \sin^2 \theta,
\eea
and hence
\bea
\frac{4}{r^2} \, F^2 - h ( G')^2 &=& 0 , \nonumber\\
 h (F')^2 -  \frac{4}{r^2} \, G^2 &=& 0 .
 \label{1a2}
 \eea
 Using
 \bea
 G' = \frac{F' \, F}{G},
 \eea
 and inserting this into the first equation of \eqref{1a2} we obtain
 \bea
 F^2 \left( \frac{4}{r^2} - h \frac{(F')^2 }{G^2}\right) = 0 \Rightarrow \frac{4}{r^2} - h \frac{(F')^2 }{G^2} = 0 ,
 \eea
 which is the second equation of \eqref{1a2}. Thus, both equations in \eqref{1a2} are consistent with one another.
It suffices to solve the second equation. Using \eqref{Ff} and \eqref{G}, we obtain
\bea
h (F')^2 -  \frac{4}{r^2} \, G^2 = 0 \Rightarrow b = \frac{a^4}{16}.
\eea
Thus,
\bea
G(r) = \sqrt{F^2(r) + \frac{a^4}{16}} = \frac{r^2}{4},
\label{G2}
\eea
and 
\bea
\rho &=& \frac14 \, r^2 \, \sqrt{h} \, \sin \theta , \nonumber\\
v &=& \frac14 \, r^2 \, \cos \theta.
\label{rvw}
\eea
We determine $e^{\psi}$ from the second equation in \eqref{drdtcond},
\bea
e^{\psi} = \frac{r^4}{16}  \, \frac{h}{F^2 \, \cos^2 \theta + G^2 \, \sin^2 \theta} = \frac{ 1 - \frac{a^4}{r^4}}{1 -  \frac{a^4}{r^4} \, \cos^2 \theta} = 
\frac{r^4 - a^4}{r^4 - a^4  \, \cos^2 \theta}.
\eea
Summarizing, we have
\bea
\Delta &=& \frac14 \, r^2 \, \left( 1 - \frac{a^4}{r^4} \right)  , \nonumber\\
B &=& \cos \theta,  \nonumber\\
e^{\psi} &=& \frac{r^4 - a^4}{r^4 - a^4  \, \cos^2 \theta}.
\label{dbp}
\eea
We still have to express $\Delta, B, e^{\psi}$ in terms of the Weyl coordinates $(\rho,v)$ given in \eqref{rvw}.
To this end we write
\bea
\rho^2 =  \frac{1}{16} (r^4 - a^4) \left(1 - 16 \frac{v^2}{r^4} \right) ,
\eea
which equals
\bea
r^8 - r^4 (a^4 + 16 (\rho^2 + v^2)) = - 16 a^4 v^2 .
\eea
This yields
\bea
r^4 = \frac12 \left[ a^4 + 16 (\rho^2 + v^2) \pm \sqrt{ (a^4 + 16 (\rho^2 + v^2) )^2 - 64 a^4 v^2} \right] .
\eea
We choose the plus sign, so as to ensure that when $a=0$, we get $r^4 = 16(\rho^2 + v^2)$,
\bea
r^4 = \frac12 \left[ a^4 + 16 (\rho^2 + v^2) + \sqrt{ (a^4 + 16 (\rho^2 + v^2) )^2 - 64 a^4 v^2} \right] .
\eea
We also have
\bea 
\cos \theta = \frac{4 v}{r^2} = \frac{4 \sqrt{2} \, v}{\sqrt{a^4 + 16 (\rho^2 + v^2) + \sqrt{ (a^4 + 16 (\rho^2 + v^2) )^2 - 64 a^4 v^2}} }.
\eea
We have thus expressed $r^2$ and $\cos \theta$ in terms of $\rho$ and $v$. Since the resulting expressions for \eqref{dbp} 
are rather long, we do not present them here.

Let us now convert back to Minkowski signature through the analytic continuation $\gamma = - i \bar t$, i.e. to 
\bea
ds^2 = - \Delta (d {\bar t} + B \, d \phi)^2 + \Delta^{-1}
\left[ e^{\psi} \left( d \rho^2 + dv^2 \right) + \rho^2 \, d \phi^2 \right], 
\label{ehmink}
\eea
with
\bea
\Delta &=& \frac14 \, r^2 \, \left( 1 - \frac{a^4}{r^4} \right) = \frac{r^4 - a^4}{4 r^2} , \nonumber\\
B &=& i \cos \theta,  \nonumber\\
e^{\psi} &=& \frac{r^4 - a^4}{r^4 - a^4  \, \cos^2 \theta}.
\label{dbpm}
\eea
Let us for convenience define the quantity $Q$ by 
\bea
Q = \frac12 \left( a^4 + 16 (\rho^2 + v^2) \right) ,
\eea
and express
\bea
r^4 = Q + \sqrt{Q^2 - 16 a^4 v^2} .
\eea
Then,
\bea
\Delta &=& \frac{  Q + \sqrt{Q^2 - 16 a^4 v^2}  - a^4}{4 \sqrt{Q + \sqrt{Q^2 - 16 a^4 v^2} }} \nonumber\\
B &=& i \frac{4 v}{\sqrt{ Q + \sqrt{Q^2 - 16 a^4 v^2} }},
\eea
and
\bea
{\tilde B} = - \frac{i}{4} \frac{Q + \sqrt{Q^2 - 16 a^4 v^2} + a^4 }{\sqrt{ Q + \sqrt{Q^2 - 16 a^4 v^2} }},
\eea
where ${\tilde B}$ is the solution to 
\bea
\star d {\tilde B} = \frac{\Delta^2}{\rho} \, dB ,
\eea
i.e. to
\bea
\partial_v {\tilde B} &=& \frac{\Delta^2}{\rho} \, \partial_{\rho} B , \nonumber\\
\partial_{\rho} {\tilde B} &=& - \frac{\Delta^2}{\rho} \, \partial_{v} B .
\eea
The matrix $M$ in \eqref{M22DB}
takes the form
\bea
M &=& 
\begin{pmatrix}
- a^4
\frac{ \sqrt{Q + \sqrt{Q^2 - 16 a^4 v^2}}}{Q + \sqrt{Q^2 - 16 a^4 v^2}  - a^4}
& \;  - i \frac{ Q + \sqrt{Q^2 - 16 a^4 v^2} + a^4}{Q + \sqrt{Q^2 - 16 a^4 v^2} - a^4}\\
- i \frac{ Q + \sqrt{Q^2 - 16 a^4 v^2} + a^4}{Q + \sqrt{Q^2 - 16 a^4 v^2} - a^4} & \;  4 \frac{ \sqrt{Q + \sqrt{Q^2 - 16 a^4 v^2}}}{Q + \sqrt{Q^2 - 16 a^4 v^2}  - a^4}
\end{pmatrix} \\
&=& \begin{pmatrix}
0  & \;  -i  \\
-i & \; 0
\end{pmatrix} + \frac{ \sqrt{Q + \sqrt{Q^2 - 16 a^4 v^2}}}{Q + \sqrt{Q^2 - 16 a^4 v^2}  - a^4} \begin{pmatrix}
- a^4& \;  0 \\
0  & \;  4 
\end{pmatrix}  \nonumber\\
&& + \frac{a^4}{Q + \sqrt{Q^2 - 16 a^4 v^2}  - a^4} \begin{pmatrix}
0 & \;  - 2 i \\
- 2 i & 0
\end{pmatrix} . \nonumber
\label{Mtt2}
\eea
Evaluating $A = M^{-1} dM$ we obtain
\bea
A = d f  \, \begin{pmatrix}
0 & 1 \\
- \frac{a^4}{4}  & 0
\end{pmatrix} 
\label{Aeha}
\eea
with 
\bea
f(\rho, v) = 4 i \frac{ \tanh^{-1} \left( \frac{\sqrt{ a^4 + 16 (\rho^2 + v^2) + \sqrt{a^8 + 32 a^4 (\rho^2 -v^2) + 256 (\rho^2 + v^2)^2 }}}{\sqrt{2} a^2} \right)}{a^2} - \frac{2 \pi }{a^2} \;,
\label{feguch}
\eea
which satisfies the harmonic equation
\eqref{eomlfa} with $\lambda =1$. The integration constant $-\frac{2 \pi }{a^2} $  was chosen such that 
when expanding $f$ in powers of $a$, the expansion starts with a zeroth order contribution given by
\bea
f_0 (\rho,v) =\frac{i}{\sqrt{\rho^2 + v^2}}
\;.
\eea
This is the function that is 
associated to the background metric, which is described by \eqref{ehmink} with $a=0$. 

Expanding $f$ in powers of $a^4$, i.e. $ f= f_0 + f_1 + {\cal O} (a^8)$, we obtain at first order in $a^4$,
\bea
f_1 (\rho, v) 
= \frac{i a^4}{96} \, \frac{2 v^2 - \rho^2}{(\rho^2 + v^2)^{5/2}} \;.
\label{dexp}
\eea
Collecting all the terms of order $a^4$ in \eqref{Aeha}, we obtain
\bea
- a^4 df_0 \, \begin{pmatrix}
0 & 0 \\
1 & 0
\end{pmatrix} + df_1 \, \, \begin{pmatrix}
0 & 1 \\
0 & 0
\end{pmatrix} \;.
\eea
Since both matrices do not commute, we have to decide which term to keep in order to obtain a class I description at order $a^4$. We discard the term proportional to $d f_0$, since it only carries information about the background. We write the term proportional to $d f_1$ as 
$df_1 \, C$, where
\bea
C = 
\begin{pmatrix}
0 & 1 \\
0  & 0
\end{pmatrix} \;.
\label{Cegh}
\eea
The function $p_1$ associated to $f_1$ is (c.f. \eqref{Pp}) 
\bea
p_1(\rho,v) &=& -  \frac{i a^4}{16} \, \frac{\rho v }{(\rho^2 + v^2)^{5/2}} \;.
\eea
Differently from all the other examples discussed in this paper, the single copy description of
the Eguchi-Hanson solution requires integrating $p_1$ \underline{twice} over $v$.
Integrating $p_1$ once with respect to $v$ (c.f. \eqref{HP}) gives
\bea
H_1(\rho, v, \phi) = \frac{i a^4}{96} \, e^{i \phi} \, \frac{\rho }{(\rho^2 + v^2)^{3/2}} \, C \;.
\eea
Integrating once more over $v$ gives 
\bea
{\hat H}_1(\rho, v, \phi) = \frac{i a^4}{96} \, e^{i \phi} \, \frac{v }{\rho \,  (\rho^2 + v^2)^{1/2}} 
 \, C \;.
\label{eghHh}
\eea

\section{Mapping gravitational solutions to solutions of the self-dual Yang-Mills equations in four dimensions 
\label{sec:grYMm}}

As already mentioned, the PDEs \eqref{emotion2d} 
that arise by dimensionally reducing gravitational theories in space-time dimensions $D \geq 4$ also arise by performing a 
suitable reduction of the self-duality equations for Yang-Mills fields in four dimensions down to two dimensions \cite{ward85,ablo,Ivanova:1994du,mas}. Using this link,
we may therefore associate solutions of the 
self-dual
Yang-Mills equations in four-dimensional flat space-time to the gravitational solutions belonging to either class I or to class II.
This is what we will do next.

We note that in these constructions 
there are two cases to consider, either $\lambda = 1$ or $\lambda =-1$. When $\lambda =1$, the flat space-time signature is $(1,3)$, whereas when $\lambda = -1$ 
the signature is $(2,2)$. 


\subsection{$\lambda =1$: signature $(1,3)$ \label{sec:ym31}}

We consider the self-dual sector of Yang-Mills (YM) theory in four-dimensional Minkowski space-time,
whose description we briefly review following \cite{lumont}, 
and perform a suitable dimensional reduction of it to two dimensions.

We introduce coordinates 
\bea
U = \tau - v \;\;\;,\;\;\; V = \tau  + v \;\;\;,\;\;\; X = - w  \;\;\;,\;\;\; 
Y = - {\bar w}  ,
\label{UVwwb}
\eea
where
\bea
w = x + i y \;\;\;,\;\;\; {\bar w} = x - i y \;,
\eea
so that
\bea 
\label{fl31}
ds^2 = - d\tau^2 + dx^2 + dy^2 + dv^2 = - dU dV + dw d {\bar w} .
\label{flat31}
\eea
We set
\bea
w = \rho \, e^{i \phi} \;.
\label{wbwrt}
\eea
We then have
\bea
\partial_U = - \frac12 \partial_v + \frac12 \partial_{\tau} \;\;\;,\;\;\; \partial_V =  \frac12 \partial_v + \frac12 \partial_{\tau} 
\label{dUV}
\eea
as well as 
\bea
\partial_w = \frac12 e^{- i \phi} \left( \partial_{\rho} - \frac{i}{\rho} \, \partial_{\phi} \right) \;\;\;,\;\;\;
\partial_{\bar w} = \frac12 e^{i \phi} \left( \partial_{\rho} + \frac{i}{\rho} \, \partial_{\phi} \right) .
\label{dwbw}
\eea
We introduce the Yang-Mills connection
\bea
\D_{\mu} = \partial_{\mu} - i g \A_{\mu} \;\;\;,\;\;\; \A_{\mu} = \A_{\mu  a} \, T_a \;\;\;,\;\;\; T_a \in \mathfrak{g} \;.
\label{covD}
\eea
Note that we are using calligraphic $\A_\mu$ to denote the genuine YM fields, in order to differentiate them from the vectors appearing in the  Breitenlohner-Maison procedure, c.f. \eqref{decompA}. Then, 
\bea
[\D_{\mu}, \D_{\nu}] = - i g  \F_{\mu \nu} 
\label{DDF}
\eea
with
\bea
\F_{\mu \nu} = \partial_{\mu} \A_{\nu} - \partial_{\nu} \A_{\mu} - i g [\A_{\mu}, \A_{\nu}] .
\eea
Self-dual solutions of the Yang-Mills equations of motion satisfy (see Appendix \ref{sec:sdymds})
\bea
\F_{\mu \nu} = \frac12 i \epsilon_{\mu \nu  \rho \sigma} \, \F^{\rho \sigma} 
\label{sdc}
\eea
with $\epsilon_{UVw\bar{w}}=\frac{i}{4}=\sqrt{-g}$.
In the coordinate system $(U,V, w, {\bar w})$, the self-duality condition \eqref{sdc} becomes (see \cite{lumont})
\bea
\F_{Uw} &=& 0 ,
\label{fuw} \\
\F_{V {\bar w} }&=& 0 ,
\label{fvbw} \\
\F_{UV} - \F_{w {\bar w}} &=& 0 .
\label{fuv}
\eea
These 3 equations are the compatibility conditions for the
following linear system (see eqs. (4) and (5) in \cite{ablo}),
\bea
L  Y = \left( \D_U   - \Omega \D_{\bar w} \right) Y = 0 \;\;\;,\;\;\; M Y =  \left( \D_w   - \Omega \D_V \right) Y  = 0.
\label{laSDYM}
\eea
Here, $Y$ is an invertible matrix function and
$\D$ denotes the covariant derivative introduced in \eqref{covD}, and $\Omega \in \mathbb{C}$ denotes a {\it constant} complex parameter,
called the spectral parameter. We take $Y \in C^2$ with respect to $(U, V, w, \bar w)$. The compatibility condition 
\bea
[ {L}, {M}] Y = 0 \;\;\; \forall \;\;\; Y \in C^2 \;\;\;,\;\;\; \Omega \in \mathbb{C} \,,
\eea
gives
\bea
0= [L, M] Y &=& \Big( [\D_U, \D_w] - \Omega \left(  [\D_U, \D_V] - [\D_w, \D_{\bar w}] \right) + \Omega^2 \, [\D_{\bar w}, \D_V] \Big) Y \nonumber\\
&=& \Big( -i \F_{Uw} + i \Omega \left( \F_{UV}- \F_{w \bar w} \right) + i \Omega^2 \, \F_{V \bar w} \Big) Y .
\label{comLM}
\eea
Then, multiplying by $Y^{-1}$ on the right gives
\bea
 -i \F_{Uw} + i \Omega \left( \F_{UV}- \F_{w \bar w} \right) + i \Omega^2 \, \F_{V \bar w}  = 0, 
 \eea
which yields \eqref{fuw}-\eqref{fuv} as vanishing coefficients of this quadratic polynomial in $\Omega$.

Note that differently from the Breitenlohner-Maison linear system
\eqref{LML}, the linear system \eqref{laSDYM} uses a constant  spectral parameter. 

Now we make the gauge choice $\A_U=0$. Then, 
\eqref{fuw} will force $\A_w =0$, subject to appropriate boundary conditions,
\bea
\A_U = \A_w =0,
\eea
in which case the linear system $(L,M)$ becomes
\bea
L   = \partial_U   - \Omega \D_{\bar w} 
\;\;\;,\;\;\; M  =   \partial_w   - \Omega \D_V \;.
\label{lax4dl1}
\eea
Note that $\A$ has been complexified; hence $\A_{\bar w}$ is not the complex conjugate of $\A_w$.

Then,   \eqref{fvbw} and \eqref{fuv} become 
\bea
0 &=& \F_{V \bar w} = \partial_V \A_{\bar w} - \partial_{\bar w} \A_V - i g [\A_V, \A_{\bar w}] ,\nonumber\\
0 &=&  \F_{UV} - \F_{w \bar w} = \partial_U \A_V - \partial_w \A_{\bar w} .
\label{int2}
\eea

Next, we restrict to self-dual solutions that are {\it static}, 
i.e. $\A_V$ and $\A_{\bar w}$ are taken to be independent of $\tau$. 
Using \eqref{dUV} and \eqref{dwbw},
the relations \eqref{int2} become
\bea
0 &=& \F_{V \bar w} = \frac12 \left(  \partial_v \A_{\bar w} - e^{i \phi} \left( \partial_{\rho} \A_V + \frac{i}{\rho} \, \partial_{\phi} \A_V \right) 
 - 2 i g [\A_V, \A_{\bar w}]  \right) ,\nonumber\\
0 &=&  \F_{UV}- \F_{w \bar w} =  - \frac12 \left(  \partial_v \A_V + e^{- i \phi} \left( \partial_{\rho} \A_{\bar w}  - \frac{i}{\rho} \, \partial_{\phi} \A_{\bar w} \right) 
\right) .
\label{int2p}
\eea
Next, let us further specialize to solutions of the form 
\bea
\A_{\bar w}  = \frac12 \,  e^{i \phi} \, A_{\rho} (\rho, v) \;\;\;,\;\;\; \A_V  = \frac12 \, A_v (\rho,v) \ ,
\label{specsol}
\eea
where the relation between the gauge algebra of the YM theory and the algebra in which the Breitenlohner-Maison vectors $A_\rho$ and $A_v$ are valued will we made explicit in Section \ref{sec:scp}. For notational simplicity, we will generally not be writing the algebra generators explicitly.
Then, \eqref{int2p} become
\bea
0 &=& \F_{V \bar w} = \frac14 e^{i \phi} \Big(  \partial_v A_{\rho} -  \partial_{\rho} A_v 
 -  i g [A_v, A_{\rho}]  \Big) ,\nonumber\\
0 &=&  \F_{UV}- \F_{w \bar w} =  - \frac14 \left(  \partial_v A_v +  \partial_{\rho} A_{\rho}  + \frac{1}{\rho} \, A_{\rho} \right) .
\label{int3}
\eea
Finally, performing the rescaling
\bea
\A \rightarrow  \frac{ i}{g} \, \A ,
\label{rescAg}
\eea
the relations \eqref{int3} become
\bea
0 &=& \F_{V \bar w} = \frac{i}{4g} e^{i \phi} \Big(  \partial_v A_{\rho} -  \partial_{\rho} A_v 
 + [A_v, A_{\rho}]  \Big) ,\nonumber\\
0 &=&  \F_{UV}- \F_{w \bar w} =  - \frac{i}{4g} \left(  \partial_v A_v +  \partial_{\rho} A_{\rho}  + \frac{1}{\rho} \, A_{\rho} \right) .
\eea
These are precisely the field equations \eqref{eoml1} and \eqref{FAA} with $\lambda =1$.

Therefore, we conclude that when $\lambda =1$, {\it any} solution to the two-dimensional PDEs \eqref{eoml1}
lifts to a static solution 
\bea
\A_U = \A_w =0 \;\;\;,\;\;\; \A_{\bar w}  = \frac12 \,  e^{i \phi} \, A_{\rho} (\rho, v) \;\;\;,\;\;\; \A_V  = \frac12 \, A_v (\rho,v) 
\label{statsol}
\eea
of the self-duality conditions for a Yang-Mills field in four flat space-time dimensions. Note that the solutions
 of \eqref{eoml1} will, in general, correspond to gravitational solutions in $D$ space-time dimensions that are not self-dual.

In the following, we specialise to gravitational solutions belonging either to class I or to class II.

\subsubsection{\bf Class I:}

Solutions in class I satisfy \eqref{decompf}. Inserting \eqref{decompf} into \eqref{statsol} gives
\bea
\A_{\bar w}  = \frac12 \,  e^{i \phi} \, \partial_{\rho} f_i (\rho, v) \, C^i \;\;\;,\;\;\; 
\A_V  = \frac12 \, \partial_v f_i(\rho,v) \, C^i 
\eea
with $C^i$ commuting matrices,  which, using \eqref{dUV} and \eqref{dwbw}, yields
\bea
\A_{\bar w}  = \partial_{\bar w } f_i (\rho, v) \, C^i \;\;\;,\;\;\; 
\A_V  =  \partial_V f_i(\rho,v) \, C^i .
\eea
Inserting this into the self-dual Yang-Mills equations \eqref{int2} gives
\bea
\left( \partial_U \partial_V - \partial_w \partial_{\bar w} \right) f_i=0 \;,
\eea
in accordance with \eqref{eomlfa}.

\subsubsection{\bf Class II:}

To a solution \eqref{ArvP} in class II, we associate the matrix function $\Psi$ defined in \eqref{PsiP}.
We may then express $A_{\rho}$ and $A_v$ as 
\bea
A_{\rho}  = - e^{-i \phi} \partial_v \Psi (\rho, v, \phi) \;\;\;,\;\;\; A_v  = e^{- i \phi} \left( \partial_{\rho} - \frac{i}{\rho} \partial_{\phi} \right) \Psi (\rho, v, \phi) \;.
\eea
Then, using \eqref{dUV} and \eqref{dwbw}, we infer that 
\bea
\A_{\bar w } = \partial_U \Psi \;\;\;,\;\;\;
\A_V = \partial_w \Psi \;\;\;.
\label{AA}
\eea
Inserting \eqref{AA} into  \eqref{int2} and performing the rescaling \eqref{rescAg} gives
\bea
\left(  \partial_U \partial_V - \partial_w \partial_{\bar w} \right) \Psi +  [\partial_w \Psi, \partial_U \Psi ] = 0 ,
\label{PE4YM}
\eea
which, defining \cite{Monteiro:2011pc,SilvLip}
\bea
\{f, g\} &=& \partial_w f \partial_U g - \partial_U f \partial_w g = \varepsilon^{\alpha \beta} \partial_{\alpha} f \partial_{\beta} g \;\;,\;\; \alpha, \beta = w, U \nonumber\\
 {[ \{f, g\} ] } &=& \varepsilon^{\alpha \beta} [ \partial_{\alpha} f,  \partial_{\beta} g ]  =  [\partial_w f, \partial_U g ] - [\partial_U f, \partial_w g ] ,
 \label{PS}
\eea
and using
\bea
[\partial_w \Psi, \partial_U \Psi ] = \frac12 [ \{\Psi, \Psi\} ],
\eea
can also be written as 
 \bea
\left( \partial_U \partial_V - \partial_w \partial_{\bar w} \right) \Psi + \frac{1}{2}  [ \{ \Psi, \Psi \} ] = 0 ,
\label{plebpsi}
\eea
which is a Plebanski type equation.

\subsection{$\lambda = -1$: signature $(2,2)$ \label{sec:22l1}}

We now change the signature of flat space-time in four dimensions from $(1,3)$ to $(2,2)$.
To this end, we consider two different coordinate systems $(U, V, X, Y) = (U, V, - w, - {\bar w})$, namely
\bea
U = \tau - v \;\;\;,\;\;\; V = \tau  + v \;\;\;,\;\;\; w =  \rho \, e^{-\phi} \;\;\;,\;\;\; {\bar w} = - \rho \, e^{\phi},
\label{UVwwbr}
\eea
and
\bea
U = \tau + i v \;\;\;,\;\;\; V = \tau - i v  \;\;\;,\;\;\; w =  \rho \, e^{i \phi} 
\;\;\;,\;\;\; {\bar w} =  \rho \, e^{-i \phi} .
\label{UVcwwb}
\eea
Then the line element $ds^2 = - dU dV + dw d {\bar w} $ becomes
\bea
 ds^2 = - d{\tau}^2 + dv^2 -  d\rho^2 + \rho^2 d\phi^2
 \label{2dvmr}
 \eea
 and
 \bea
 ds^2  = - d{\tau}^2 - dv^2 +  d\rho^2 + \rho^2 d\phi^2 \,,
 \label{3dvmr}
 \eea
respectively. Note that the associated two-dimensional line element is $dv^2 - d\rho^2$ in the first case, and $- dv^2 + d\rho^2$ in the second case. The Yang-Mills self-duality condition \eqref{sdc} again takes the form given in \eqref{fuw}-\eqref{fuv} in both these coordinate systems (see Appendix \ref{sec:sdymds}).

Let us first consider the coordinate system \eqref{UVwwbr}, in which case
\bea
&&\partial_U = - \frac12 \partial_v + \frac12 \partial_{\tau} \;\;\;,\;\;\; \partial_V =  \frac12 \partial_v + \frac12 \partial_{\tau} \;,  \nonumber\\
&&\partial_w = \frac12 e^{\phi} \left( \partial_{\rho} - \frac{1}{\rho} \, \partial_{\phi} \right) \;\;\;,\;\;\;
\partial_{\bar w} = - \frac12 e^{-\phi} \left( \partial_{\rho} + \frac{1}{\rho} \, \partial_{\phi} \right) .
\eea
Similar to the discussion given above for signature $(1,3)$, 
we specialise to {\it static} solutions, of the form
\bea
\A_U = \A_w = 0 \;\;\;,\;\;\; 
 \A_{\bar w}  =  - \frac12 \, e^{-\phi} \, A_{\rho}  (\rho, v) \;\;\;,\;\;\;
 \A_V  = \frac12 A_v (\rho, v) \;.
\label{c1}
\eea
Then, inserting this into the self-dual Yang-Mills equations \eqref{int2} gives, upon performing the rescaling \eqref{rescAg},
\bea
0 &=& \F_{V \bar w} \Longrightarrow
-\frac14 \, e^{ -\phi}  \Big( \partial_v A_{\rho} - \partial_{\rho} A_v + [A_v, A_{\rho}] \Big) =0 \;, \nonumber\\
 0 &=&  \F_{UV}- \F_{w \bar w} = 
 \partial_U \A_V- \partial_{w} \A_{\bar w} \Longrightarrow  - \frac14 \left( \partial_v A_v - \partial_{\rho} A_{\rho} - \frac{A_{\rho}}{\rho} \right) = 0 .
 \eea
These are precisely the field equations \eqref{eoml1} and \eqref{FAA} with $\lambda = -1$.

Next, let us consider the coordinate system \eqref{UVcwwb}, in which case
 \bea
 \partial_U &=& - \frac{i}{2} \partial_v + \frac12 \partial_{\tau} \;\;\;,\;\;\;
  \partial_V =  \frac{i}{2} \partial_v + \frac12 \partial_{\tau} , \nonumber\\
\partial_w &=& \frac12 e^{- i \phi} \left( \partial_{\rho} - \frac{i}{\rho} \, \partial_{\phi} \right) \;\;\;,\;\;\;
\partial_{\bar w} = \frac12 e^{i \phi} \left( \partial_{\rho} + \frac{i}{\rho} \, \partial_{\phi} \right) .
\label{derc2}
\eea
We again specialise to {\it static solutions}, of the form
\bea
\A_U = \A_w = 0 \;\;\;,\;\;\; \A_{\bar w}  =   \frac12 \, e^{i\phi} \, A_{\rho}  (\rho, v) \;\;\;,\;\;\; \A_V  = \frac{i}{2} A_v (\rho, v) \;.
\label{c2}
\eea
Then, inserting this into the self-dual Yang-Mills equations \eqref{int2} gives, upon performing the rescaling \eqref{rescAg},
\bea
0 &=& \F_{V \bar w} \Longrightarrow
\frac{i}{4} \, e^{ i\phi}  \Big( \partial_v A_{\rho} - \partial_{\rho} A_v + [A_v, A_{\rho}] \Big) =0 \;, \nonumber\\
 0 &=&  \F_{UV}- \F_{w \bar w} = 
 \partial_U \A_V- \partial_{w} \A_{\bar w} \Longrightarrow  \frac14 \left( \partial_v A_v - \partial_{\rho} A_{\rho} - \frac{A_{\rho}}{\rho} \right) = 0 .
 \eea
Again, these are precisely the field equations \eqref{eoml1} and \eqref{FAA} with $\lambda = -1$.

Finally, we note that we could also have started from the $(2,2)$ line element 
\bea
ds^2 = dU dV - dw d{\bar w}
\eea
instead. Then, using the coordinate system \eqref{UVcwwb}, we get
\bea
 ds^2  =  d{\tau}^2 + dv^2 -  d\rho^2 - \rho^2 d\phi^2 \,.
 \label{3dvmrkl}
 \eea
 The self-duality condition \eqref{sdc}  with $\epsilon_{\tau\rho v\phi} = \sqrt{-g}$ yields 
the set of equations \eqref{fuw}-\eqref{fuv}.
 Then, specialising to static solutions of the form \eqref{c2} results again in the field equations \eqref{eoml1} and \eqref{FAA} with $\lambda = -1$.

In the following, we specialise to gravitational solutions belonging either to class I or to class II.

\subsubsection{\bf Class I:}

Solutions in class I satisfy \eqref{decompf}. Inserting \eqref{decompf} into either \eqref{c1} or \eqref{c2}
gives
\bea
\A_{\bar w} (\rho, v, \phi) = \partial_{\bar w } f_i (\rho, v) \, T^i \;\;\;,\;\;\; 
\A_V (\rho, v, \phi) =  \partial_V f_i(\rho,v) \, T^i .
\eea
Inserting this into the self-dual Yang-Mills equations \eqref{int2} gives
\bea
\left( \partial_U \partial_V - \partial_w \partial_{\bar w} \right) f_i=0 \;,
\eea
in accordance with \eqref{eomlfa}.

\subsection{\bf Class II:}

Let us consider solutions of the form \eqref{c2}. Using \eqref{ArvP} 
as well as 
$\Psi$ defined in \eqref{PsiP}, we obtain
\bea
\A_{\bar w } = i \partial_U \Psi \;\;\;,\;\;\;
\A_V = i \partial_w \Psi \;\;\;,
\label{AA2}
\eea
where we used \eqref{derc2}.
Inserting \eqref{AA2} into  \eqref{int2} and performing the rescaling \eqref{rescAg} gives
\bea
\left( \partial_U \partial_V - \partial_w \partial_{\bar w} \right) \Psi +i [\partial_w \Psi, \partial_U \Psi ] = 0 ,
\eea
which, defining \cite{Monteiro:2011pc,SilvLip}
\bea
\{f, g\} &=& \partial_w f \partial_U g - \partial_U f \partial_w g = \varepsilon^{\alpha \beta} \partial_{\alpha} f \partial_{\beta} g \;\;,\;\; \alpha, \beta = w, U \nonumber\\
 {[ \{f, g\} ] } &=& \varepsilon^{\alpha \beta} [ \partial_{\alpha} f,  \partial_{\beta} g ]  =  [\partial_w f, \partial_U g ] - [\partial_U f, \partial_w g ] ,
\eea
and using
\bea
[\partial_w \Psi, \partial_U \Psi ] = \frac12 [ \{\Psi, \Psi\} ],
\eea
can also be written as 
 \bea
\left(  \partial_U \partial_V - \partial_w \partial_{\bar w} \right) \Psi + \frac{i}{2}  [ \{ \Psi, \Psi \} ] = 0 .
\eea

On the other hand, for solutions of the form \eqref{c1},
for which the coordinates $(U,V, w, {\bar w})$ are all real, we define
$\Psi$ to be
\bea
\Psi (\rho, v, \phi) = - \frac12 \, e^{-\phi} \, P(\rho,v) \;.
\label{PsiReal}
\eea
Then, using \eqref{ArvP}, together with \eqref{PsiReal} and \eqref{c1}, we have
\bea 
\A_{\bar w}  =-\partial_U \Psi \;\;\;,\;\;\;
\A_V  = -\partial_w \Psi \;.
\eea
Inserting these into \eqref{int2} and performing the rescaling \eqref{rescAg}, we get 
\bea
\left( \partial_U \partial_V   - \partial_{w} \partial_{\bar w}\right) \Psi - \frac12 [\{ \Psi, \Psi\}]=0 \;.
\eea

\section{Single copy prescription \label{sec:scp}}

 Let us consider gravitational solutions in four dimensions
 that admit both a class I and a class II description.
 We remark that one  could set up the single copy prescription using only a class I description, but additionally putting the solution in class II allows us to directly recover the scalar version of the self-dual Yang-Mills solution, c.f.
\eqref{genPiAmu}.

Let us assume that these solutions depend on certain parameters, and that when switching off these parameters the solution 
describes a flat space-time solution, with all other fields 
vanishing. This, however, does not necessarily imply that the gravitational 1-form $A = M^{-1} d M $ 
vanishes when switching off these parameters. An example thereof is provided by the Eguchi-Hanson solution. We will therefore have to distinguish between two cases: either $A$ vanishes or it does not vanish when switching
off these parameters. In the following, we will first consider the case when $A$ vanishes in this limit.
The other case will be discussed in Section \ref{sec:egAnv}.

Let us consider working to first order in one of the parameters, which we call $m$. We assume that when $m$ vanishes,  $A$ also vanishes. Then the 
gravitational 1-form $A = M^{-1} d M $ is of order $m$, whereas the field $\psi$ in the line element \eqref{4dWLP} is order $m^2$
in view of \eqref{eq_psi}. Therefore, at first order in $m$, all the information about the perturbed solution is encoded in the gravitational 1-form $A = M^{-1} d M $.

 Let us assume that to first order in $m$, the gravitational 1-form $A$  takes the form
\bea
A = df_i \, C_i \;,
\label{Adfcg}
\eea
where the $C_i$ are constant commuting matrices and where the 
 $f_i = f_i(\rho,v)$ satisfy \eqref{eomlfa}. 
Further, let us assume that 
$A$ can be brought to the form \eqref{ArvP}, with $P$ given by $P(\rho, v) = p_i(\rho,v) \, C_i$
and satisfying \eqref{harmt},
\bea
A_{\rho} = - \frac{\lambda}{2} \partial_v p_i (\rho, v) \, C_i\;\;\;,\;\;\; A_{v} = \frac12 \left( \partial_{\rho} + \frac{1}{\rho} \right) p_i (\rho, v)  \, C_i \;.
\eea
This is the case in all the examples discussed in this paper, see Section \ref{sec:exam12}.

 To such a gravitational solution we would now like to associate a solution of the self-dual Yang-Mills equations, as described 
in the previous section, which is the single copy of \eqref{Adfcg}. Since the associated Yang-Mills solution is self-dual, whereas the gravitational solution is in general not, to match with the single copy literature we will have to discard half of the content of the former to obtain the single copy of
the gravitational solution.

We will first consider the case $\lambda =1$ and subsequently discuss the case $\lambda =-1$.

\subsection{$\lambda =1$}

We introduce the coordinates
$(U,V,w,{\bar w})$ given in \eqref{UVwwb}. In these coordinates, we introduce the operators\footnote{The operators $\hat\Pi$ have also appeared in \cite{SilvLip,Campiglia:2021srh,CarrilloGonzalez:2024sto,Nagy:2022xxs}.}
\bea
({\hat \Pi}_U, {\hat \Pi}_V, {\hat \Pi}_w, 
{\hat \Pi}_{\bar w} ) = ( 0, \partial_w, 0, \partial_U) \;,
\eea
which satisfy
\bea
{\hat \Pi}\cdot {\hat \Pi} = 0 \;\;\;,\;\;\; {\hat \Pi} \cdot \partial = 0 \;,
\eea
where the inner product is taken with respect to the flat metric
in \eqref{flat31}. Then, we may associate to
\eqref{Adfcg} the following static  Yang-Mills solution (c.f. \eqref{AA})
\bea\label{genPiAmu}
\A_{\mu} = {\hat \Pi}_{\mu} \Psi \;\;\;,\;\;\; \mu= U, V, w, {\bar w} \;,
\eea
where $\Psi$ is determined in terms of $P$ via \eqref{PsiP}.

Now recall that given a solution \eqref{Adfcg}, we may
construct a solution $A = dG_i \, C_i$, where the functions $G_i$ are constructed as in 
\eqref{fftG} by integration over $v$, and a solution $H = H_i \, C_i$,
where the functions $H_i$  are constructed as in 
\eqref{HP} by integration over $v$.
Continuing this procedure, we may 
in principle integrate $G_i$ over $v$ to produce yet another solution, and so forth. To each of these gravitational solutions 
we then may associate a Yang-Mills solution as described above.
Which of these Yang-Mills solutions will correspond to the single copy associated with the gravitational solution
\eqref{Adfcg}?

Our proposal for the single copy associated with the gravitational solution \eqref{Adfcg} is as follows. 
Rather than starting from the gravitational solution \eqref{Adfcg}, we instead start from the associated 
gravitational solution $A = (dG_i) \, C_i$. This is motivated by physical considerations. Even though we have not presented the sources explicitly, we have used them as a guide for a physically reasonable single copy: the guiding principle is that we would like to map sources of the same type, i.e. monopole to monopole, dipole to dipole etc. 

We construct the static Yang-Mills solution in two steps. In the first step we write the vectors

\be \label{singcop0}
A_U = A_w =0 \;,
\ee
and for class I
\be 
\begin{aligned}
A_{\bar w}  &=
\frac12 \,  e^{i \phi} \, \partial_{\rho} G_i(\rho,v) \, C_i 
\;, \\
A_V  &= \frac12 \, \partial_v G_i (\rho,v) \, C_i 
\;,
\label{sc1}
\end{aligned}
\ee
alternatively for class II
\be 
\begin{aligned}
A_{\bar w}  &=& 
-\frac12 \partial_v H
&=& 
 \partial_U H   \;, \\
A_V  &=& 
\frac12 \, e^{- i \phi} \, \left( \partial_{\rho} + \frac{1}{\rho} \right) H  &=& \partial_w H 
\;.
\label{sc}
\end{aligned}
\ee
Note that \eqref{sc} is automatically of the form \eqref{genPiAmu}.

In the above, to get to \eqref{sc1}, we used the uplift \eqref{statsol}. One can map from class I to class II by an application of \eqref{relsGH}, and this gives the first equality in \eqref{sc}. The second equality in \eqref{sc} is reached via the coordinate change \eqref{dUV}, \eqref{dwbw}.  
We note that, using \eqref{relsGH}, \eqref{Pp} and  $f_i=\partial_v G_i$ we can alternatively write the non-vanishing gauge fields in terms of the original functions describing class I and II solutions, as :
\be \label{singcop2}
\begin{aligned}  
A_{\bar w} &=- \frac14 \, e^{i \phi} \, p_i(\rho,v) \, C_i \;, \\
A_V&=\frac12 \, f_i (\rho,v) C_i \;.
\end{aligned} 
\ee 
For the non-vanishing components of the field strength we obtain, making use of \eqref{sc}, \eqref{singcop2} and \eqref{relsGH}
\bea
F_{UV} = F_{w \bar w} &=& - \frac12 \partial_v A_V = 
- \frac14 \partial_v f_i \, C_i  = -\frac18 \left( \partial_{\rho} + \frac{1}{\rho} \right) p_i \, C_i  \,, \nonumber\\
F_{U \bar w} &=& - \frac12 \partial_v A_{\bar w}  =
- \frac14 e^{i \phi} \partial_{\rho} f_i \, C_i = \frac18 e^{i \phi} \, \partial_v p_i \, C_i   \;, \nonumber\\
F_{V w} &=& - \partial_w A_V = - \frac14 e^{-i \phi} \partial_{\rho} f_i \, C_i = \frac18 e^{-i \phi} \, \partial_v p_i \, C_i \;.
\label{fsI}
\eea
Note that the field strength components are given in terms of single derivatives of $f_i$ (or $p_i$).
We now proceed to the second step, and promote the vectors in \eqref{singcop0} and \eqref{sc}, to proper Yang-Mills (YM) vectors via
\be \label{CitoTa}
A_\mu=A_{\mu i}C_i\quad\to\quad \mathcal{A}_{\mu}=\mathcal{A}_{\mu a}T_a \;,
\ee 
where $T_a$ are the generators of some non-abelian gauge algebra $\mathfrak{g}$, such that $C_i$ are equal to a commuting subset of the $T_a$'s. We see that, unlike in the usual single copy story, this places certain restrictions of the type and size of the YM gauge group. This will turn out to have interesting consequences, as we will detail below. We can make the map in \eqref{CitoTa} precise via 
\be \label{sc_map_genrs}
\A_{\mu a}=
\left\{\begin{matrix}
A_{\mu i}, & a=i \\ 
0, & \text{otherwise}
\end{matrix}\right.
\ee 
Thus we have `enlarged' the matrix-valued vectors in \eqref{singcop0} and \eqref{sc} to proper Yang-Mills vectors.

As an aside, we note that under the procedure above, we also have
\be 
H=H_i C_i\quad\to\quad \mathcal{H}=\mathcal{H}_a T_a
\ee 
with $\mathcal{H}_a$ constructed from $H_i$ as in \eqref{sc_map_genrs}. We then see that the role of $\Psi$ in \eqref{genPiAmu} is played by our integrated function $H_i$, and thus we can think of $H_i$ as a `seed' for both the double and 0th copy procedure, in the self-dual sector:
\be 
\mathcal{H}^{(0)}\ \xleftarrow[]{\text{0th copy}}\ \mathcal{A}_\mu= {\hat \Pi}_{\mu}\mathcal{H}\ \xrightarrow[]{\text{double copy}}\ 
h_{\mu\nu}={\hat \Pi}_{\mu}{\hat \Pi}_{\nu}\mathcal{H}^{(2)} \;,
\ee 
where $\mathcal{H}^{(2)}$ is a function, and $\mathcal{H}^{(0)}=\mathcal{H}^{(0)}_{a\tilde{a}}T_a\tilde{T}_{\tilde{a}}$ is a bi-adjoint scalar.

All the examples of gravitational solutions with $\lambda =1$ discussed in Section \ref{sec:exam12} have both a class I and a class II description,
and to each of them we associated functions $G_i$ and $H_i$.
We may therefore associate a single copy to each of these gravitational solutions using \eqref{sc} and \eqref{CitoTa}.
As an illustrative example, let us consider 
the example in \eqref{fp}, based on a single function $f$, for which we obtain using \eqref{UVwwb},
\bea\label{singlecopyschwarzschild1}
\F_{\tau v a} &=& 2 \F_{UV a} = - m \,  \frac{v}{(\rho^2 + v^2)^{3/2}} \, T_a  \,, \nonumber\\
\F_{\tau  x a} & = & \F_{U \bar w a} + \F_{V w a} = - m \, \cos \phi \, \frac{\rho}{(\rho^2 + v^2)^{3/2}} \, T_a \,, \nonumber\\
\F_{\tau y a} &=& - i \left( \F_{U \bar w a} - \F_{V w a} \right) = 
- m \, \sin \phi \, \frac{\rho}{(\rho^2 + v^2)^{3/2}} \, T_a
\eea
which, when converting to spherical coordinates 
using 
$v = r \cos \theta, \, \rho = r \sin \theta$, gives
\bea\label{singlecopyschwarzschild2}
\F_{\tau v a} &=& - m \, \frac{ \cos \theta}{r^2} \, T_a \,, \nonumber\\
\F_{\tau x a}&=& - m \, \frac{ \cos \phi \, \sin \theta }{r^2}  \, T_a \,, \nonumber\\
\F_{\tau y a}&=& - m \, \frac{ \sin \phi \, \sin \theta }{r^2}  \, T_a \,,
\eea
which describes the electric field produced by a point charge. Since the solution is self-dual, there is 
also an associated magnetic monopole field. If we discard the magnetic monopole solution,
then this describes the single copy associated to the gravitational solution \eqref{fp} discussed in \cite{Monteiro:2014cda}.

\subsubsection{The nontrivial role of the gauge group}

 Let us compare our single copy prescription with what is done in the context of the double copy. 
In the double copy framework, a gravitational solution is constructed as a ``product" of two gauge theory solutions. Depending on the complexity of the gravity solution, the two gauge theory factors might be identical (e.g. for Schwarzschild, in which case we have a unique single copy) or different (this is often the case for gravity solutions supported by additional fields).  For example the black hole in Section \ref{Black hole in ADS2XS2} is supported by a Maxwell field. Thus it will arise from an asymmetric double copy, where one of the factors is a pure Yang-Mills theory, and the other is Yang-Mills coupled with a scalar\footnote{For some explicit examples of asymmetric double copies for black hole solutions see \cite{Cardoso:2016amd,Cardoso:2016ngt}.}. 
 Generally in the double copy procedure, the gauge algebra $\mathfrak{g}$ is unspecified, and independent of the specific kinematics or profile of the solutions considered. 
 
 Our single copy proposal is drastically different,
 in that we have a unique single copy for which 
  the (non-compact) Yang-Mills gauge group is determined by the Breitenlohner-Maison procedure itself. Thus various aspects of the gravitational solution will be encoded on the Yang-Mills side in the Yang-Mills gauge group generators supporting the solution.

Let us summarise a few interesting cases below, all of which share the kinematics described in \eqref{singlecopyschwarzschild1} and \eqref{singlecopyschwarzschild2}, i.e. they are all of the form
\be 
A = df \, C
\ee 
with the same 
\be\label{common_f}
f =  - 2 \frac{m}{\sqrt{\rho^2 + v^2}} \ ,
\ee
but with distinct gauge groups (i.e. different $C$'s), encoding additional information about the gravity solution.

\begin{itemize}
\item Exterior region of the Schwarzschild black hole, linearised in $m$ (see Section \ref{subsec:Exterior region of the Schwarzschild black hole} for details). The Breitenlohner-Maison procedure dictates that the vector field is valued in $G/H = SL(2, \mathbb{R})/SO(2)$. The matrix $M$ is given by 
\be 
M^{(S)}=\left(
\begin{array}{cc}
 1+f & 0 \\
 0 & 1-f \\
\end{array}
\right) \;.
\ee 
Then for the single copy procedure we will construct $A=M^{-1}dM=df C$, and we will  take the gauge group to be 
\be 
G_{YM}^{(S)}= SL(2, \mathbb{R}) \;.
\ee 
We have a single non-vanishing component in \eqref{sc_map_genrs} with the corresponding generator given by
 \be 
 T_1^{(S)}\equiv C=\begin{pmatrix} 
1 & 0 \\ 0 & -1
\end{pmatrix} \;.
 \ee 
Then, taking into account \eqref{singcop0} and \eqref{sc}, the YM gauge field is, explicitly 
\be\label{explicit_components}
\begin{aligned} 
\A_U =&0 \,,\\
\A_V  =& \frac12 f\ T^{(S)}_1 \,,\\
\A_w =&0 \,,\\
\A_{\bar w}  =& 
\frac12 \,  e^{i \phi} \, \partial_{\rho} G\ T^{(S)}_1 \,,
\end{aligned}
\ee 
with $f$ as \eqref{common_f}, and $G$ obtained by integration from $f$ as in \eqref{fftG} and \eqref{gp} 
(recall that the coordinates $(U, V, w , \bar w)$ are related to the coordinates $(\rho, v, \phi)$ through \eqref{UVwwb} and \eqref{wbwrt}).

 \item Black hole in $AdS_2 \times S^2$ (see Section \ref{Black hole in ADS2XS2}). This can be obtained from the Schwarzschild solution in two steps. First we embed $M^{(S)}$ into $G/H = SU(2,1) / ( SL(2, \mathbb{R} ) \times U(1))$ via 
\be 
M^{(emb)}=
\left(
\begin{array}{ccc}
 1-f & 0 & 0 \\
 0 & 1 & 0 \\
 0 & 0 & 1+f \\
\end{array}
\right) \;.
\ee 
Then, we obtain the matrix $M$ for a black hole in $AdS_2 \times S^2$, namely
\be
{\tilde M} = g^{\natural} ({c}) M g({ c}) \;,
\ee
where
 \bea
 g(c) = \begin{pmatrix}
 1 & 0 & 0 \\
 - \sqrt{2} {\bar c}  & 1 & 0 \\
 |c|^2 &  - \sqrt{2} c & 1
 \end{pmatrix} \;\;\;,\;\;\; 
 c \in \mathbb{C} \;,
 \eea
and $ g^{\natural} (c) = \eta g(c)^{\dagger} \eta^{-1}$ with $\eta = {\rm diag} (1, -1, 1)$. 
Then we have
\be
\tilde{M}=
\left(
\begin{array}{ccc}
 \left(|c|^2-1\right)^2 -f\left(1 - |c|^4\right) & \sqrt{2} c \left(1-|c|^2 \left(1 + f\right)\right) & |c|^2 \left(1 + f\right) \\
 \sqrt{2} {\bar c} \left(| c| ^2\left(1 + f\right)-1\right) & 1-2|c|^2 \left(1 + f\right) & \sqrt{2} {\bar c} \left(1 + f\right) \\
 |c|^2 \left(1 + f\right) & -\sqrt{2} c \left(1+ f\right) & 1+ f \\
\end{array}
\right) \;.
\ee
Again we construct $A=\tilde M^{-1}d\tilde M=df \tilde C$, obtaining
\be 
\tilde{C}= -
 \begin{pmatrix} 
 1 & 0 & 0\\
 \sqrt{2} {\bar c} & 0 & 0 \\
 0 & \sqrt{2} c & -1
 \end{pmatrix}
\ee
and we will  take the gauge group to be 
\be 
G_{YM}^{(AdS_2 \times S^2)}= SU(2,1) \;.
\ee 
We have
\be \tilde{C}
= - H_2 -  \sqrt{2} \left( {\rm Re} \, c \right)\, (F_1 + F_2) + \sqrt{2} \left( {\rm Im} \, c \right) \, i(F_2 - F_1) \;,
\ee
where $H_2, F_1 + F_2, i(F_2 -F_1)$ denote a subset of generators of $SU(2,1)$ given in (A.6) and (A.12) in \cite{Camara:2017hez}. Then we identify our gauge group via
\be 
T_1^{(AdS_2 \times S^2)}\equiv  H_2,\qquad
T_2^{(AdS_2 \times S^2)}\equiv F_1 + F_2, \qquad
T_3^{(AdS_2 \times S^2)}\equiv i(F_2 - F_1)
 \ee 
and the YM gauge field components are given by
\be \label{explicit_components2}
\begin{aligned} 
\A_U =&0 \,,\\
\A_V  =& \frac12 f\left( - T_1^{(AdS_2 \times S^2)} - \sqrt{2} \left( {\rm Re} \, c \right)\, T_2^{(AdS_2 \times S^2)} + \sqrt{2} \left( {\rm Im} \, c \right) \, T_3^{(AdS_2 \times S^2)} \right) \,,\\
\A_w =&0 \,,\\
\A_{\bar w}  =& 
\frac12 \,  e^{i \phi} \, \partial_{\rho} G \left( - T_1^{(AdS_2 \times S^2)} -  \sqrt{2} \left( {\rm Re} \, c \right)\, T_2^{(AdS_2 \times S^2)} + \sqrt{2} \left( {\rm Im} \, c \right) \, T_3^{(AdS_2 \times S^2)} \right)\,,
\end{aligned}
\ee 
where $f$ and $G$ are the same as in \eqref{explicit_components}. Thus we already see that the gravitational information is not lost when taking the single copy, but rather repackaged in the matrices associated with generators of the gauge group. 

Finally, we remark that the YM generators associated to the black hole in $AdS_2 \times S^2$  can be constructed via a global non-abelian transformation generated by the very same matrix that gave us the Harrison transformation. Explicitly, the embedded 1-form $A$ is given by
\be 
A^{(emb)}=\left[M^{(emb)}\right]^{-1}dM^{(emb)}=df C^{(emb)}, \quad C^{(emb)}= - \begin{pmatrix} 
 1 & 0 & 0\\
 0 & 0 & 0 \\
 0 & 0 & -1
 \end{pmatrix} 
\ee 
and indeed the gauge field is obtained via
\be 
\mathcal{A}_\mu^{(AdS_2 \times S^2)}=g(c)^{-1}\mathcal{A}_\mu^{(emb)}g(c) \ .
\ee
\\

 \item Self-dual Taub-NUT (see Section \ref{sec:Self-dual Taub-NUT solution}).
The matrix $M(\rho,v)$ belonging to a complexification of $G/H = SL(2, \mathbb{R})/SO(2)$ is given by
\be 
M^{(TN)}= \begin{pmatrix}
- (\alpha^2 -1) + (\alpha-1)^2 f
& i \alpha +i (1-\alpha) f\\
  i \alpha +i (1-\alpha) f
  &1-f
 \end{pmatrix} \; \;\;,\;\;\; \alpha \in \mathbb{C} \;.
\ee 
We have $A^{(TN)} = \left[M^{(TN)}\right]^{-1}d M^{(TN)} = df \, C^{(TN)}$, with 
\bea
C^{(TN)} = \begin{pmatrix}
1- \alpha  & i \\
i (1-\alpha)^2 & \alpha -1
\end{pmatrix} \;.
\eea
We then pick the gauge group to be
\be 
G_{YM}^{(TN)}=SL(2,\mathbb{C}) \;.
\ee
A set of Lie algebra generators for $SL(2,\mathbb{C})$ is given by
\be \label{subset_gen_SDTN}
T_1^{(TN)}=
\begin{pmatrix}
1 & 0 \\
0 & -1
\end{pmatrix},\quad
T_2^{(TN)}=
\begin{pmatrix}
0 & i \\
0 & 0
\end{pmatrix},\quad
T_3^{(TN)}=
\begin{pmatrix}
0 & 0 \\
i & 0
\end{pmatrix} \;,
\ee 
such that
\be 
C^{(TN)}=\left(1 - \alpha \right)T_1^{(TN)}
+T_2^{(TN)}+\left(1-\alpha\right)^2 T_3^{(TN)}
\ee 
and thus we have  
\be 
\begin{aligned} 
\A_U =&0 \,,\\
\A_V  =& \frac12 f\left(\left(1-\alpha \right)T_1^{(TN)}
+T_2^{(TN)}+\left(1-\alpha\right)^2 T_3^{(TN)}\right) \,,\\
\A_w =&0 \,,\\
\A_{\bar w}  =& 
\frac12 \,  e^{i \phi} \, \partial_{\rho} G\left(\left(1-\alpha\right)T_1^{(TN)}
+T_2^{(TN)}+\left(1-\alpha\right)^2 T_3^{(TN)} \right) \ ,
\end{aligned}
\ee 
again with the same $f$ and $G$ as in \eqref{explicit_components} and \eqref{explicit_components2}.
\end{itemize}

\subsubsection{Solutions with multiple generators}

An example of the form \eqref{Adfcg} based on two commuting matrices $C_1$ and $C_2$ is provided by the Kaluza-Klein black hole described in Section \ref{subsec: Kaluza-Klein black hole: rotating attractor}. Its single copy description is then based on two functions $G_1$ and $G_2$ (respectively $H_1$ and $H_2$) with
\bea
\A_V = \frac12 \left( f_1(\rho,v) \, C_1 + f_2 (\rho, v) \, C_2 \right)\,,
\eea
where $f_1$ and $f_2$ are given in \eqref{KKf1f2}.
\bea
C_1 = \begin{pmatrix}
0 & 0 & 0 \\
- \frac{{\cal B}}{{\cal D}} & 0 & 0 \\
0 & \frac{{\cal B}}{{\cal C}} & 0 
\end{pmatrix} \;\;\;,\;\;\; C_2 = \begin{pmatrix}
0 & 0 & 0 \\
0 & 0 & 0 \\
- \frac{ 2 {\cal A D} + {\cal B}^2 }{2 {\cal C D} } & 0 & 0
\end{pmatrix} \;.
\eea
Note that this example does not require an expansion in a parameter $m$, since the field $e^{\psi}$ in the line element \eqref{4dWLP} satisfies $e^{\psi}=1$ \cite{Camara:2017hez}. The coset is $G/H = SL(3,\mathbb{R})/SO(2,1)$, so following the prescription from the previous subsections, we define
\be 
G_{YM}^{(KK)}=SL(3,\mathbb{R}) \,.
\ee 
The following subset of Lie algebra generators of $SL(3,\mathbb{R})$ will be relevant
\be 
T_1^{(KK)}=\begin{pmatrix}
0 &0  &0 \\ 
1 &0  &0 \\ 
0 &0  &0 
\end{pmatrix}\ ,\quad
T_2^{(KK)}=\begin{pmatrix}
0 &0  &0 \\ 
0 &0  &0 \\ 
0 &-1  &0 
\end{pmatrix} \ , \quad
T_3^{(KK)}=\begin{pmatrix}
0 &0  &0 \\ 
0 &0  &0 \\ 
1 &0  &0 
\end{pmatrix} \,.
\ee 
We can write $C_1$ and $C_2$ in terms of the above as
\be 
C_1=- \frac{{\cal B}}{{\cal D}} T_1^{(KK)}
-\frac{{\cal B}}{{\cal C}}T_2^{(KK)}, \quad
C_2=- \frac{ 2 {\cal A D} + {\cal B}^2 }{2 {\cal C D} } T_3^{(KK)}
\ee 
and then the components of the gauge field are given explicitly as 
\be 
\begin{aligned} 
\A_U =&0 \,,\\
\A_V  =& -\frac12 f_1\left(\frac{{\cal B}}{{\cal D}} T_1^{(KK)}
+\frac{{\cal B}}{{\cal C}}T_2^{(KK)}\right)
-\frac12 f_2\frac{ 2 {\cal A D} + {\cal B}^2 }{2 {\cal C D} } T_3^{(KK)} \,,\\
\A_w =&0 \,,\\
\A_{\bar w}  =& 
-\frac12 \,  e^{i \phi} \, \partial_{\rho} G_1\left(\frac{{\cal B}}{{\cal D}} T_1^{(KK)}
+\frac{{\cal B}}{{\cal C}}T_2^{(KK)}\right) 
-\frac12 \,  e^{i \phi} \, \partial_{\rho} G_2 
\frac{ 2 {\cal A D} + {\cal B}^2 }{2 {\cal C D} } T_3^{(KK)} \,,
\end{aligned}
\ee 
with $f_1$, $f_2$ given in \eqref{KKf1f2} and $G_1$, $G_2$ given in \eqref{KKG1G2}.

\subsection{$\lambda = - 1$}
For $\lambda = -1$, the procedure is completely analogous to the one for $\lambda=1$. In particular \eqref{singcop0} remains the same, whereas \eqref{sc} and \eqref{fsI}  need to be slightly modified to account for the new coordinate systems. We have two possibilities: we may consider either the coordinate system \eqref{UVwwbr} or \eqref{UVcwwb}. Let us first consider the latter. Then, using \eqref{c2}, the first step in our single copy prescription is 
\bea
A_U &=& A_w = 0 \;, \nonumber\\
A_{\bar w}  &=&
 \partial_{\bar w} G_i \, C_i =
\frac12 \,  e^{i \phi} \, \partial_{\rho} G_i(\rho,v) \, C_i = 
i \partial_U H = 
\frac14 \, e^{i \phi} \, p_i(\rho,v) \, C_i  \;, \nonumber\\
A_V  &=& \partial_V G_i \, C_i = \frac{i}{2} \, f_i (\rho,v) \, 
C_i =  
i \partial_w H = \frac{i}{2} \, e^{- i \phi} \, \left( \partial_{\rho} + \frac{1}{\rho} \right) H \;.
\eea
For the non-vanishing components of the field strength we obtain
\bea
F_{UV} &=& F_{w \bar w} = - \frac{i}{2} \partial_v A_V = 
\frac14 \partial_v f_i \, C_i  = \frac18 \left( \partial_{\rho} + \frac{1}{\rho} \right) p_i \, C_i  \,, \nonumber\\
F_{U \bar w} &=& - \frac{i}{2} \partial_v A_{\bar w}  =
- \frac{i}{4} e^{i \phi} \partial_{\rho} f_i \, C_i = - \frac{i}{8} e^{i \phi} \, \partial_v p_i \, C_i   \;, \nonumber\\
F_{V w} &=& - \partial_w A_V = - \frac{i}{4} e^{-i \phi} \partial_{\rho} f_i \, C_i = - \frac{i}{8} e^{-i \phi} \, \partial_v p_i \, C_i \;.
\eea
Note that the field strength components are given in terms of single derivatives of $f_i$ (or $p_i$). In the second step we promote the above to YM fields, following the same procedure as for $\lambda=1$, see eqn. \eqref{CitoTa} and \eqref{sc_map_genrs}.

Using this prescription, we may associate a single copy to the Einstein-Rosen wave solution \eqref{ersol}. The Breitenlohner-Maison procedure dictates that the vector field is valued in $G/H = SL(2, \mathbb{R})/SO(2)$. The matrix $M$ is given by 
\be 
M^{(ER)}=\left(
\begin{array}{cc}
 e^f & 0 \\
 0 & e^{-f} \\
\end{array}
\right)
\ee 
with 
\bea
f(\rho,v) = 2 \cos (k  v ) 
\, J_0(k \rho) -2 \;\;\;,\;\; C = \begin{pmatrix}
1 & 0 \\ 0 & -1
\end{pmatrix} \;.
\eea
Then for the single copy procedure we will construct $A=M^{-1}dM=df C$, and we will  take the gauge group to be 
\be 
G_{YM}^{(ER)}= SL(2, \mathbb{R}) \;.
\ee 
We have a single non-vanishing component in \eqref{sc_map_genrs}, with the corresponding generator given by
 \be 
 T_1^{(ER)}\equiv C=\begin{pmatrix}
1 & 0 \\ 0 & -1
\end{pmatrix} \,,
 \ee 
i.e. the YM gauge field is, explicitly 
\be
\begin{aligned} 
\A_U =&0 \,, \\
\A_V  =& \frac{i}{2} f\ T^{(ER)}_1 \,, \\
\A_w =&0 \,, \\
\A_{\bar w}  =& 
\frac12 \,  e^{i \phi} \, \partial_{\rho} G\ T^{(ER)}_1 \,.
\end{aligned}
\ee 
To lowest order in $k$, $f$ and $G$ are given by \eqref{fGk2}.

As a further example, let us consider 
the Kleinian Taub-NUT solution\footnote{A single copy description of this Kleinian space-time 
has also been discussed recently in 
\cite{Easson:2023dbk,Desai:2024fgr} using a different approach based on Kerr-Schild coordinates.} described in Section \ref{sec:Self-dual Kleinian Taub-NUT solution}. The coset is a complexification of $G/H = SL(2, \mathbb{R})/SO(2)$, and the Breitenlohner-Maison vector is given by
\bea
A = df \, C \; ,\;\ f(\rho,v) = - \frac{2 m}{\sqrt{v^2 - \rho^2}}  \;,\; C = \begin{pmatrix}
1+\alpha & -i \\
-i (1+\alpha)^2 & -(1+\alpha)
\end{pmatrix} \;,\; C^2 = 0 \:.
\eea
We take the gauge group to be
\be 
G_{YM}^{(KTN)}=SL(2,\mathbb{C}) \;.
\ee
Let us use the same subset of Lie algebra generators as in 
\eqref{subset_gen_SDTN}:
\be 
T_1^{(KTN)}=
\begin{pmatrix}
1 & 0 \\
0 & -1
\end{pmatrix},\quad
T_2^{(KTN)}=
\begin{pmatrix}
0 & i \\
0 & 0
\end{pmatrix},\quad
T_3^{(KTN)}=
\begin{pmatrix}
0 & 0 \\
i & 0
\end{pmatrix} \;.
\ee 
Then we have
\be 
C=(1+\alpha)T_1^{(KTN)} -T_2^{(KTN)} -(1+\alpha)^2 T_3^{(KTN)} \,, 
\ee 
so that the YM field components are given by
\be 
\begin{aligned} 
\A_U =&0 \,,  \\
\A_V  =& \frac{i}{2} f\left((1+\alpha)T_1^{(KTN)} -T_2^{(KTN)} -(1+\alpha)^2T_3^{(KTN)}\right) \,, \\
\A_w =&0 \,,  \\
\A_{\bar w}  =& 
\frac12 \,  e^{i\phi} \, \partial_{\rho} G\left((1+\alpha)T_1^{(KTN)} -T_2^{(KTN)} -(1+\alpha)^2T_3^{(KTN)}\right) \, .
\end{aligned}
\ee 

Finally, let us also consider the coordinate system \eqref{UVwwbr}.
 Then, using \eqref{c1}, the first step in our single copy prescription is
\bea
A_U &=& A_w = 0 \;, \nonumber\\
A_{\bar w}  &=&
 \partial_{\bar w} G_i \, C_i =
- \frac12 \,  e^{- \phi} \, \partial_{\rho} G_i(\rho,v) \, C_i = 
- \partial_U H = 
- \frac14 \, e^{- \phi} \, p_i(\rho,v) \, C_i  \;, \nonumber\\
A_V  &=& \partial_V G_i \, C_i = \frac{1}{2} \, f_i (\rho,v) 
\, C_i =  
- \partial_w H = -  \frac{1}{2} \, e^{  \phi} \, \left( \partial_{\rho} + \frac{1}{\rho} \right) H \;,
\eea
where we used
\bea 
A_{\bar w}  =-\partial_U H
\;\;\;,\;\;\;
A_V  = -\partial_w H
\eea
with 
\bea
H(\rho, v, \phi) = - e^{- \phi}  \left( \frac12 \int^{v}_a P(\rho, {\tilde v}) \, d {\tilde v} +  h(\rho) \right) \;\;\;,\;\;
P(\rho, {v}) = p_i (\rho,v) \, C_i \;.
\label{HP2}
\eea
For the non-vanishing components of the field strength we obtain
\bea
F_{UV} &=& F_{w \bar w} = - \frac{1}{2} \partial_v A_V = 
- \frac14 \partial_v f_i \, C_i  = - \frac18 \left( \partial_{\rho} + \frac{1}{\rho} \right) p_i \, C_i  \,, \nonumber\\
F_{U \bar w} &=& - \frac{1}{2} \partial_v A_{\bar w}  =
\frac{1}{4} e^{ -\phi} \partial_{\rho} f_i \, C_i =  \frac{1}{8} e^{- \phi} \, \partial_v p_i \, C_i   \;, \nonumber\\
F_{V w} &=& - \partial_w A_V = - \frac{1}{4} e^{ \phi} \partial_{\rho} f_i \, C_i = - \frac{1}{8} e^{ \phi} \, \partial_v p_i \, C_i \;.
\label{fsI2}
\eea
In the second step we promote the above to YM fields, following the same procedure as for $\lambda=1$, see eqn. \eqref{CitoTa} and \eqref{sc_map_genrs}.

\subsection{The case of the Eguchi-Hanson metric \label{sec:egAnv}}

In this subsection we consider the Eguchi-Hanson metric \eqref{ehmink}. 
As already remarked, when switching off the parameter $a$, the associated matrix 1-form $A$ 
does not vanish (c.f. \eqref{Aeha}), even though the metric is flat when written in lightcone coordinates,
see \eqref{cartegha0}. This differs from all the other examples considered in this paper.
To deal with this situation, we 
expand the matrix 1-form $A$ in powers of the parameter $a^4$, 
\bea
A = A^{(0)} + A^{(1)} + \dots \,.
\eea
We call $A^{(0)}$ the background 1-form, while
$A^{(1)}$ denotes its perturbation at first order in the perturbation parameter $a^4$.

Now we focus on $ A^{(1)} $ and invoke the construction given in Section \ref{sec:2D4D}.
We express $ A^{(1)}$ in terms of the four-dimensional lightcone coordinates $(U, V, X, Y)$ given below, and 
subsequently we focus on the component set $(0, A^{(1)}_V , 0, A^{(1)}_Y ) $. We use 
the four-dimensional lightcone
coordinates 
introduced in \cite{Berman:2018hwd}. Performing the analytic continuation $\gamma=  - i \bar t$, we obtain
\bea
U &=& r \, \sin \frac{\theta}{2} \, e^{\frac{i}{2} \,  \phi - \frac12 \bar t} \;, 
\nonumber\\
V &=& - r \, \sin \frac{\theta}{2} \, e^{- \frac{i}{2} \,  \phi + \frac12 \bar t } \;, \nonumber\\
X &=& - w =  - r \, \cos \frac{\theta}{2} \, e^{\frac{i}{2} \,  \phi + \frac12 \bar t} \;, \nonumber\\
Y &=& - {\bar w } = - r \, \cos \frac{\theta}{2} \, e^{- \frac{i}{2} \,  \phi - \frac12 \bar t} \;,
\eea
which satisfy
\bea
r^2 = X Y - U V \;.
\eea
In these lightcone coordinates, and to first order in the parameter $a^4$, the Eguchi-Hanson metric \eqref{ehmink} 
takes the form
\bea
ds^2_4 = - d U dV + dX dY + \frac{a^4}{2} \frac{  \left( V dU - X dY \right)^2 + \left( U dV - Y d X \right)^2 }{ \left( X Y - U V \right)^3 } \;.
\label{cartegha0}
\eea

The component set $(0, A^{(1)}_V , 0, A^{(1)}_Y ) $ is expressed in terms of a matrix function 
(c.f. \eqref{Apotpsps}), which in this example is the matrix function ${\hat H}_1$ given in \eqref{eghHh},
\bea
A^{(1)}_V  = 
- \partial_X {\hat H}_1 \;\;\;,\;\;\;
A^{(1)}_Y
= - 
\partial_U {\hat H}_1 \;.
\eea
Using
\bea
 e^{2i \phi } = - \frac{U \, X }{V \, Y} \Rightarrow  e^{i \phi } = \sqrt{ - \frac{U \, X }{V \, Y} }
 \eea
as well as
\bea
\sqrt{\rho^2 + v^2} &=& \frac14 \left( X Y - U V  \right) \;, \nonumber\\
\rho &=& \frac12 \, \sqrt{- X Y U V } \;, \nonumber\\
v &=& \frac14 \left( X Y + U V \right) \:,
\eea
we get
\bea
{\hat H}_1(U, V, X, Y) 
= \frac{i a^4}{48} \, \left( \frac{1}{Y V } +  \frac{2}{X Y - U V} \, \frac{U}{Y} \right) \, C \;.
\eea
The first term, proportional to $1/(Y V)$ gets projected out when computing
$A^{(1)}_V, A^{(1)}_Y$,
\bea
A^{(1)}_V  &=& 
- \partial_X {\hat H}_1 = \frac{i a^4}{24} \frac{U}{(X Y - U V)^2} \; C \;, \nonumber\\
A^{(1)}_Y 
&=& -
\partial_U {\hat H}_1 = - \frac{i a^4}{24}  \frac{X}{(X Y - U V)^2} \; C \;.
\eea
The associated matrix field strength $F^{(1)}$ reads
\bea
F^{(1)} &=& \frac{i a^4}{24 (U V - X Y )^3} \\
&& \left[ - (U V + X Y ) \left( d U \wedge d V + d X \wedge d Y \right)
+ 2 V X dU \wedge d Y + 2 U Y d X \wedge d V \right] \, C \;. \nonumber
\eea
This is the expression for the field strength is the Lorentzian space-time \eqref{cartegha0}.
To compare with the expressions for the gauge field and its field strength in Euclidean space with line element $ds_4^2 = d U dV - dX dY$,
given in eq. (5.9) in \cite{Luna:2018dpt} and in 
eq. (114) in 
\cite{Alawadhi:2019urr} respectively, we first interchange $U$ and $V$ in $F^{(1)}$, which corresponds to passing from the self-dual to the anti self-dual sector, c.f. Appendix \ref{sec:asdymds}. Subsequently we 
send $U \rightarrow i U, V \rightarrow i V , X \rightarrow i X, Y \rightarrow i Y$. The resulting expressions for $A^{(1)}$ and $F^{(1)}$ agree with the expressions given  in \cite{Luna:2018dpt,Alawadhi:2019urr}.
The field strength $F^{(1)}$ can be promoted to a field strength of a non-abelian Yang-Mills gauge group as in \eqref{CitoTa}.


\section{Conclusions \label{sec:conc}}

In this paper we discussed the mapping of the integrable sector of certain gravitational theories in $D \geq 4$ to the integrable sector of Yang-Mills theories in four dimensions. This mapping is formulated in terms of Weyl coordinates, since these are the natural coordinates used in the Lax pair description of the integrable sector of these gravitational theories. The gauge group on the Yang-Mills side is specified by the dimensional reduction 
performed on the gravitational side. Let us summarize our results.

We considered two classes of solutions (class I and class II) to the two-dimensional PDEs \eqref{grfe2}.
These PDEs arise by dimensional reduction of the field equations of certain gravitational theories in $D \geq 4$ dimensions, and therefore solutions belonging to these two classes correspond to solutions of the gravitational field equations in $D$ space-time dimensions. The two-dimensional PDEs
\eqref{grfe2} are the compatibility conditions for a Lax pair, the Breitenlohner-
Maison linear system. This Lax pair can be described by a pair of differential operators $(\cal L, \cal M)$ in two dimensions,
which have been given in Section \ref{sec:dimred}. 

The PDEs \eqref{grfe2} also arise as compatibility conditions for certain Lax pairs in three and in four dimensions, see Appendix \eqref{sec:laxLM}. That there is a Lax pair in four dimensions giving rise to \eqref{grfe2} is not surprising, since it is
known \cite{Ivanova:1994du,mas} that a certain dimensional reduction of the self-duality equations for Yang-Mills
fields in four dimensions gives rise to the two-dimensional PDEs \eqref{grfe2}. Solutions to the PDEs \eqref{grfe2} can thus be regarded as either describing gravitational solutions in $D \geq 4$ or 
describing solutions of the self-dual Yang-Mills equations in four dimensions.

The single copy prescription associates a Yang-Mills solution to a gravitational solution. To ensure that this mapping is physically meaningful, we demand 
that the nature of the sources of both types of solutions be the same. Our single copy prescription is as follows.
We focus on gravitational solutions that have both a class I and a class II description.
These gravitational solutions are constructed from the 1-form $A$ which we assume to depend on a parameter
which, when set to zero, 
gives rise to a background solution $A^{(0)}$. 
Subsequently we work at first order in this parameter.
By expanding $A$ in powers of this parameter, $A = A^{(0)} + A^{(1)} + \dots $,  we obtain the first-order perturbation $A^{(1)}$. 
 Then, given $A^{(1)}$,
we construct a new gravitational solution by integrating over the Weyl coordinate $v$, as depicted in Table \ref{table:Ansig} for the case $\lambda = 1$ and in Table \ref{table:Ansig2} for the case $\lambda = -1$. 
We then use this new solution as a device 
for constructing the single copy $\A$ of the original gravitational solution, that is, for ensuring that the nature of the sources supporting the gravitational and the Yang-Mills solutions is the same (monopole to monopole, etc.). We illustrated our single copy construction in various examples and confirmed that it correctly reproduces the single copy for various gravitational solutions discussed in the literature. In all but one of the examples the background $A^{(0)}$ vanishes and we integrated once over $v$.
The exception is the Eguchi-Hanson solution, which we discussed in Section \ref{sec:egAnv}:
its background $A^{(0)}$ is non-vanishing, and the single copy requires integrating twice over $v$.

\begin{table}
\begin{center}
\begin{tabular}{|c | c | c | c|}
 \hline
 $\lambda =1$ & $A^{(1)}$ & 
 & signature $(1,3)$: \, $\A$  \\ [0.5ex] 
 \hline\hline
 Class I & $d f \, C $ & $G(\rho, v) = \int^{v}_a f(\rho, {\tilde v}) \, d {\tilde v}  + g(\rho) $  & 
 \makecell{$\A_U = \A_w = 0$, \\  $\A_{\bar w} = \partial_{\bar w} G \, C, \, \A_V = \partial_V G \, C$ }\\ 
 \hline 
 Class II & $ - \frac{1}{2}\, \star d P + \frac{P}{2 \rho} \, dv  $ & 
 $ H(\rho, v, \phi) = \frac12 e^{i \phi}  \left(\int^{v}_a P(\rho, {\tilde v}) \, d {\tilde v} +  h(\rho) \right)   $ & \makecell{ $\A_U = \A_w = 0$, \\ $ \A_{\bar w} = \partial_U H, \, \A_V = \partial_w H $ }\\ 
 \hline
 \end{tabular}
 \caption{Single copy construction for $\lambda = 1$. The coordinates $(U, V, w, \bar w) $ are given in \eqref{UVwwb} and \eqref{wbwrt}. The Hodge star $\star$ is defined in \eqref{stho}.}
 \label{table:Ansig}
\end{center}
\end{table}

\begin{table}
\begin{center}
\begin{tabular}{|c | c | c | c |}
 \hline
$\lambda = -1$   & $A^{(1)}$ & 
& signature $(2,2)$: \, $\A$  \\ [0.5ex] 
 \hline\hline
 Class I & $d f \, C $ & $G(\rho, v) = \int^{v}_a f(\rho, {\tilde v}) \, d {\tilde v}  + g(\rho) $  & \makecell{$\A_U = \A_w = 0$, \\ $\A_{\bar w} = \partial_{\bar w} G \, C, \, \A_V = \partial_V G \, C$  }\\ 
 \hline 
 Class II & $ \frac{1}{2}\, \star d P + \frac{P}{2 \rho} \, dv  $ & 
 $ H(\rho, v, \phi) = \frac12 e^{i \phi}  \left(\int^{v}_a P(\rho, {\tilde v}) \, d {\tilde v} +  h(\rho) \right)   $  & \makecell{$\A_U = \A_w = 0$, \\ $\A_{\bar w} =i \partial_U H, \, \A_V = i\partial_w H $ }\\  [0.5ex] 
 \hline
 Class II & $ \frac{1}{2}\, \star d P + \frac{P}{2 \rho} \, dv  $ 
 & $H(\rho, v, \phi) = -e^{ -\phi}  \left( \frac12 \int^{v}_a P(\rho, {\tilde v}) \, d {\tilde v} +  h(\rho) \right) $ & \makecell{$\A_U = \A_w = 0$, \\ $\A_{\bar w} =- \partial_U H, \, \A_V = - \partial_w H $ }\\ [0.5ex] 
 \hline
 \end{tabular}
 \caption{Single copy construction for $\lambda = -1$. The coordinates $(U, V, w, \bar w )$ are given in \eqref{UVcwwb} and in \eqref{UVwwbr}. The Hodge star $\star$ is defined in \eqref{stho}.}
 \label{table:Ansig2}
\end{center}
\end{table}


\section*{Acknowledgements}
We would like to thank Cameron Beetar, David Berman and Thomas Mohaupt for discussions and Suresh Nampuri for very helpful 
discussions and comments on the draft.
The authors would like to thank the Isaac Newton Institute for Mathematical Sciences, Cambridge, for support and hospitality during the programme {\it Black holes: bridges between number theory and holographic quantum information}, where work on this paper was undertaken. This work was supported by EPSRC grant EP/R014604/1. GLC would like to thank Durham University for hospitality during the course of this work.
GLC was partially supported
by FCT/Portugal through CAMGSD, IST-ID, projects UIDB/04459/2020 and UIDP/04459/2020. S.N. is supported in part by STFC consolidated grant T000708.

\appendix

\section{Various Lax pairs \label{sec:laxLM}}

The PDEs \eqref{emotion2d} can be viewed as being the compatibility conditions for a Lax pair of differential operators. Here we discuss 4 such pairs. 

The first Lax pair $(L,M)$, formulated in $D=4$ in coordinates $(U, V, w, {\bar w})$, is given in \eqref{lax4dl1}, 
\bea
L   = \partial_U   - \Omega \left( \partial_{\bar w} + \A_{\bar w} \right) 
\;\;\;,\;\;\; M  =   \partial_w   + \Omega \left( \partial_V + \A_V \right) \;,
\eea
with $\A_{\bar w}, \A_V$ expressed in terms of $A_{\rho}, A_v$ as in \eqref{statsol} when $\lambda =1$, 
and as in \eqref{c1} and in \eqref{c2} when $\lambda =-1$. Here, $\Omega$ is a constant spectral parameter.

The second Lax pair $({\hat L}, {\hat M}) $ is formulated in $D=3$ in coordinates $(\rho, v, \phi)$, and again uses a constant spectral parameter $\Omega$,
\bea
{\hat L} &=& - e^{-2 i \phi} \, \partial_v + \frac12 \rho \, \Omega \,  \left(  D_{\rho} + \frac{i}{\rho} \, \partial_{\phi}  \right) \;, \nonumber\\
 {\hat M} &=&   e^{- i \phi} \, \lambda \, \partial_{\rho} + \frac12 \rho \, \Omega \, e^{ i \phi} \,   D_{v} \;\;\;,\;\;\; \lambda = \pm 1,
\eea
where the covariant derivatives $D$ are given by
\bea
D_{\rho} &=& \partial_{\rho} +   A_{\rho},  \nonumber\\
D_v &=& \partial_v + A_v , \nonumber\\
 { [D_{\rho}, D_v] }&=& F_{\rho v} .
 \label{covld3}
\eea
Note the presence of the factor $\rho$ multiplying the constant spectral parameter $\Omega$.
The compatibility condition
\bea
{\hat L} Y = 0 \; \wedge \; {\hat M} Y = 0 \Longrightarrow [{\hat L}, {\hat M}] Y = 0 \quad \forall \quad Y \in C^2 \;\;,\;\; \Omega \in \mathbb{C} 
\eea
gives
\bea
0 = [{\hat L}, {\hat M}] Y =  \left[ - \frac12 \rho \, \Omega \, e^{- i \phi}  \left(  \partial_v A_v + \lambda \, \partial_{\rho} A_{\rho} + \frac{\lambda }{\rho} A_{\rho} 
 \right)  +  \left(  \frac12 \rho \, \Omega \right)^2 \, e^{i \phi}\, F_{\rho v}\right] Y .
\eea
Taking $Y$ to be an invertible matrix function and multiplying by $Y^{-1}$ on the right gives
 \eqref{eoml1} and \eqref{FAA} as vanishing coefficients of this quadratic polynomial in $\Omega$.

The third Lax pair $({\check L}, {\check M}) $ is also formulated in $D=3$ in coordinates $(\rho, v, \phi)$, 
but now it uses the non-constant spectral parameter $\tau$ specified by \eqref{relt2}, which is 
a function of $\rho, v$ and of $\omega \in \mathbb{C}$ and satisfies
\bea
\partial_v \tau &=& \frac{2 \lambda \tau^2}{\rho (\tau^2 + \lambda)} ,\nonumber\\
\partial_{\rho} \tau &=& \frac{\tau (\lambda - \tau^2)}{ \rho(  \tau^2 + \lambda)} .
\label{dertspec}
\eea
The Lax pair $({\check L}, {\check M}) $ reads
\bea
{\check L}&=& e^{ i \phi} \left( -  \partial_v + \tau  \,  \left(  D_{\rho} + \frac{i}{\rho} \, \partial_{\phi}  \right) \right) \;, \nonumber\\
 {\check M} &=&   e^{- i \phi} \left(  \lambda \, \partial_{\rho} +  \tau \,   D_{v} \right) \;\;\;,\;\;\; \lambda = \pm 1,
 \label{3dhlax}
\eea
where the covariant derivatives $D$ are given by \eqref{covld3}.
The compatibility condition
\bea
{\check L} Y = 0 \; \wedge \; {\check M} Y = 0 \Longrightarrow [{\check L}, {\check M}] Y = 0 \quad \forall \quad Y \in C^2 \;\;,\;\; \Omega \in \mathbb{C} 
\eea
yields \eqref{eoml1} and \eqref{FAA}, as follows. We compute
\bea
 [{\check L}, {\check M}]&=&  - \tau \left( \partial_v A_v + \lambda \partial_{\rho} A_{\rho} +  \frac{\lambda}{\rho}  A_{\rho} \right) + \tau^2 F_{\rho v} \nonumber\\
&& + \frac{\tau}{\rho } \left( \lambda D_{\rho} + \tau D_v \right) 
 - \partial_v \tau \, D_v - \lambda \partial_{\rho} \tau  \left(  D_{\rho} + \frac{i}{\rho} \, \partial_{\phi}  \right) \nonumber\\
&& + \tau \partial_{\rho} \tau \, D_v - \tau \partial_v \tau  \left(  D_{\rho} + \frac{i}{\rho} \, \partial_{\phi}  \right) + i \frac{\lambda \tau}{\rho^2} \, \partial_{\phi} .
\eea
Using \eqref{dertspec}, we find that all the terms in the second and third lines cancel out,
resulting in 
\bea
0 = [{\check L}, {\check M}] Y = \left[  - \tau \left( \partial_v A_v + \lambda \partial_{\rho} A_{\rho} + \frac{\lambda}{\rho} A_{\rho} \right) + \tau^2 F_{\rho v} \right] Y  .
\eea
Assuming that the matrix function $Y$ is invertible 
and multiplying by $Y^{-1}$ on the right gives, using $\tau \neq 0$,
 \bea
  -  \left(  \partial_v A_v + \lambda \, \partial_{\rho} A_{\rho} + \frac{\lambda }{\rho} A_{\rho} 
 \right)  +  \tau \, F_{\rho v}  = 0 .
 \label{sumedp}
 \eea
 Now we note that $F_{\rho v}$ has to vanish, because otherwise we would obtain
 \bea
 \tau = - \frac{ \left(  \partial_v A_v + \lambda \, \partial_{\rho} A_{\rho} + \frac{\lambda }{\rho} A_{\rho} 
 \right) }{ F_{\rho v}},
 \eea
 which cannot hold, since the right hand side is independent of $\omega$, whereas $\tau$ is a function of $\omega$. Hence we infer that \eqref{eoml1} and \eqref{FAA} hold.


The fourth Lax pair $({\cal L},{\cal M}) $, formulated in $D=2$ in coordinates $(\rho,v)$, is given in \eqref{LML},
\bea
\mathcal{L} &=& - \partial_v + \tau \, D_{\rho} 
= - \partial_v + \tau \, \left( \partial_{\rho} + A_{\rho} \right)  , \nonumber\\
 \mathcal{M} &=& 
 \lambda \, \partial_{\rho} + \tau \, D_v
= 
 \lambda \, \partial_{\rho} + \tau \, \left( \partial_v + A_v \right)  \;\;\;,\;\;\; \lambda = \pm 1, 
 \label{LMLex}
 \eea
where $\tau$ is a non-constant spectral parameter, namely a function of $\rho, v$ and of $\omega \in \mathbb{C}$ as specified by \eqref{relt2}.
The compatibility condition 
\bea
{\mathcal L} X = 0 \; \wedge \; {\mathcal M} X = 0 \Longrightarrow [ {\mathcal L}, {\mathcal M}] X = 0 \quad \forall \quad X \in C^2 \;\;,\;\; \omega \in \mathbb{C} 
\eea
yields \eqref{eoml1} and \eqref{FAA}, as follows. Using
\bea
[\mathcal{L}, \mathcal{M}] = -  \left( [\partial_v, \tau \, D_v] + \lambda \, [\partial_{\rho}, \tau D_{\rho} ]\right) +  [\tau \, D_{\rho}, \tau \, D_v] 
\eea
as well as
\bea
 [D_{\rho}, D_v] = F_{\rho v} \;\;\;,\;\;\; 
 { [\partial_v  , D_v ] } = \partial_v A_v \;\;\;,\;\;\;   [\partial_{\rho}  , D_{\rho} ] = \partial_{\rho} A_{\rho} , 
   \eea
we obtain
\bea
0 = [{\cal L}, {\cal M}] X= - \frac{\lambda}{\tau} \left( - \partial_v \tau + \tau \partial_{\rho} \tau \right) \partial_{\rho}
- \left( \lambda \partial_{\rho} \tau + \tau \partial_v \tau \right) D_{\rho} -\tau \left( \partial_v A_v + \lambda \partial_{\rho} A_{ \rho} \right) + \tau^2 \, F_{\rho v},
\label{cor1}
\eea
where we used ${\cal M} X = 0$ once. Next, using the relations
\eqref{dertspec} in the form
\bea
 - \partial_v \tau + \tau \partial_{\rho} \tau &=& - \frac{\tau^2}{ \rho}  \;,\nonumber\\
  \lambda \partial_{\rho}\tau + \tau \partial_v \tau &=& \lambda \frac{\tau}{\rho}  \;,
 \eea
 and inserting them into \eqref{cor1} gives
 \bea
0 = [{\cal L}, {\cal M}] X= \left[ 
-\tau \left( \partial_v A_v + \lambda \partial_{\rho} A_{ \rho}  + \frac{\lambda}{ \rho}  \, A_{\rho} \right)  + \tau^2 \, F_{\rho v}
\right] X.
\label{cor2}
\eea
Assuming that the matrix function $X$ is invertible 
and multiplying by $X^{-1}$ on the right gives, using $\tau \neq 0$,
 \bea
  -  \left(  \partial_v A_v + \lambda \, \partial_{\rho} A_{\rho} + \frac{\lambda }{\rho} A_{\rho} 
 \right)  +  \tau \, F_{\rho v}  = 0 .
 \eea
By an argument analogous to the one used below \eqref{sumedp}, we conclude that \eqref{eoml1} and \eqref{FAA} both hold.

\section{Self-dual Yang-Mills equations in different space-time signatures \label{sec:sdymds}}

\subsection{Signature $(1,3)$:}

We consider the line element in signature $(1,3)$,
\bea
ds^2 = - dx_0^2 + dx_1^2 + dx_2^2 + dx_3^2 \;.
\eea
Given a two-form $\F$, the self-duality condition $\F = \star \F$ implies the relations 
\bea
\F_{01} = i \F_{23} \;,\; \F_{02} = -i \F_{13} \;,\; \F_{03} = i \F_{12} \;.
\label{sdrel13}
\eea
Setting 
\bea
x_0 = \tau \;,\; x_1 = v \;,\; x_2 = \rho \, \cos \phi \;,\; x_3 = \rho \, \sin \phi \;,
\eea
and introducing the coordinates
\bea
U = \tau -  v \;\;\;,\;\;\; V = \tau +  v  \;\;\;,\;\;\; w =  \rho \, e^{i \phi} 
\;\;\;,\;\;\; {\bar w} =  \rho \, e^{-i \phi} ,
\eea
we obtain the line element \eqref{flat31},
\bea
 ds^2  = - d{\tau}^2 + dv^2 +  d\rho^2 + \rho^2 d\phi^2 =  - dU dV + dw d {\bar w} \;.
  \eea
Using
\bea
dx_0 \wedge dx_1 &=& \frac12 dU \wedge d V \;, \nonumber\\
dx_2 \wedge dx_3 &=&
\rho \,  d \rho\, \wedge d \phi = \frac{i}{2} d w \wedge d {\bar w} \;, 
\eea
we obtain
\bea
 \F_{01} \, dx_0 \wedge dx_1 &=& \frac12 \F_{01} \, dU \wedge d V = \F_{U V } \, dU \wedge d V \;, \nonumber\\
 \F_{23} \, dx_2 \wedge dx_3 &=& \frac{i}{2} \F_{23}\, d w \wedge d {\bar w} = 
\F_{w \bar w} \, d w \wedge d {\bar w} \;, 
\eea
and we infer the relations 
\bea
\F_{U V } = \frac12 \F_{01} \;\;\;,\;\;\; \F_{w \bar w} =  \frac{i}{2} \F_{23} \;.
\eea
The, using \eqref{sdrel13}, we obtain
\bea
\F_{UV} =  \frac12 \F_{01} \ = \frac{i}{2} \F_{23} = \F_{w {\bar w}} \;,
\eea
in agreement with \eqref{fuv}. Similarly, one infers the relations \eqref{fuw} and \eqref{fvbw}.

\subsection{Signature $(2,2)$:}

We consider the line element in signature $(2,2)$,
\bea
ds^2 = dx_0^2 + dx_1^2 - dx_2^2 - dx_3^2 \;.
\eea
Given a two-form $\F$, the self-duality condition $\F = \star \F$ implies the relations \cite{Hamanaka:2020pyc}
\bea
\F_{01} = \F_{23} \;,\; \F_{02} = \F_{13} \;,\; \F_{03} = - \F_{12} \;.
\label{sdrel22}
\eea

Now let us perform various coordinate changes. First, consider 
setting 
\bea
x_0 = \rho \, \cos \phi \;,\; x_1 = \rho \, \sin \phi \;,\; x_2 = \tau \;,\; x_3 = v \;,
\eea
in which case we obtain the line element \eqref{3dvmr},
\bea
 ds^2  = - d{\tau}^2 - dv^2 +  d\rho^2 + \rho^2 d\phi^2 \,.
 \label{mtmvrrp}
  \eea
Then, introducing the coordinates
\bea
U = \tau + i v \;\;\;,\;\;\; V = \tau - i v  \;\;\;,\;\;\; w =  \rho \, e^{i \phi} 
\;\;\;,\;\;\; {\bar w} =  \rho \, e^{-i \phi} ,
\eea
the line element becomes expressed as 
\bea
ds^2 = - dU dV + dw d {\bar w} \;.
\eea
Using
\bea
dx_0 \wedge dx_1 &=& \rho \,  d \rho\, \wedge d \phi = \frac{i}{2} d w \wedge d {\bar w} \;, \nonumber\\
dx_2 \wedge dx_3 &=& \frac{i}{2} dU \wedge d V \;,
\eea
we obtain
\bea
 \F_{01} \, dx_0 \wedge dx_1 &=& \frac{i}{2} \F_{01} \, d w \wedge d {\bar w} = 
\F_{w \bar w} \, d w \wedge d {\bar w} \;, \nonumber\\
 \F_{23} \, dx_2 \wedge dx_3 &=& \frac{i}{2} \F_{23} \, dU \wedge d V = \F_{U V } \, dU \wedge d V ,
\eea
and we infer the relations 
\bea
\F_{U V } =  \frac{i}{2} \F_{23} \;\;\;,\;\;\; \F_{w \bar w} =  \frac{i}{2} \F_{01} \;.
\eea
The, using \eqref{sdrel22}, we obtain
\bea
\F_{UV} =  \frac{i}{2}  \F_{23} \ = \frac{i}{2} \F_{01} = \F_{w {\bar w}} \;,
\label{fuv2}
\eea
in agreement with \eqref{fuv}. Similarly, one infers the relations \eqref{fuw} and \eqref{fvbw}.

Next, let us consider setting 
\bea
x_0 = \rho \, \sinh \phi \;,\;  x_1 = v \;,\;  x_2 = \tau \;,\;
x_3 = \rho \, \cosh \phi \;,
\eea
in which case we obtain the line element \eqref{2dvmr},
\bea
 ds^2 = - d{\tau}^2 + dv^2 -  d\rho^2 + \rho^2 d\phi^2 \;.
  \eea
Then, introducing the coordinates
\bea
U = \tau - v \;\;\;,\;\;\; V = \tau +  v  \;\;\;,\;\;\; w =  \rho \, e^{- \phi} 
\;\;\;,\;\;\; {\bar w} = - \rho \, e^{ \phi} ,
\eea
the line element becomes expressed as 
\bea
ds^2 = - dU dV + dw d {\bar w} \;.
\eea
Using
\bea
dx_0 \wedge dx_3 &=&  - \rho \,  d \rho\, \wedge d \phi =  \frac{1}{2} d w \wedge d {\bar w} \;, \nonumber\\
dx_1 \wedge dx_2 &=& - \frac{1}{2} dU \wedge d V \;,
\eea
we obtain
\bea
 \F_{03} \, dx_0 \wedge dx_3 &=&  \frac12 \F_{03} \, d w \wedge d {\bar w} = 
\F_{w \bar w} \, d w \wedge d {\bar w} \;, \nonumber\\
 \F_{12} \, dx_1 \wedge dx_2 &=& - \frac12  \F_{12} \, dU \wedge d V = \F_{U V } \, dU \wedge d V ,
\eea
and we infer the relations 
\bea
\F_{U V } = -\frac12 \F_{12} \;\;\;,\;\;\; \F_{w \bar w} =  \frac12 \F_{03} \;.
\eea
Then, using \eqref{sdrel22}, we obtain
\bea
\F_{UV} = - \frac12  \F_{12} \ =  \frac12 \F_{03} = \F_{w {\bar w}} \;,
\eea
in agreement with \eqref{fuv}. Similarly, one infers the relations \eqref{fuw} and \eqref{fvbw}.

Finally, we note that the line element \eqref{3dvmrkl} differs from \eqref{mtmvrrp} by an overall sign, 
and hence the relation \eqref{fuv2} applies to this case as well.

\section{Anti self-dual Yang-Mills equations in signature $(1,3)$ \label{sec:asdymds}}

We consider the line element in signature $(1,3)$,
\bea
ds^2 = - dx_0^2 + dx_1^2 + dx_2^2 + dx_3^2 \;.
\eea
Given a two-form $\F$, the anti self-duality condition $\F = - \star \F$ implies the relations 
\bea
\F_{01} = - i \F_{23} \;,\; \F_{02} = i \F_{13} \;,\; \F_{03} = - i \F_{12} \;.
\label{asdrel13}
\eea
Setting 
\bea
x_0 = \tau \;,\; x_1 = v \;,\; x_2 = \rho \, \cos \phi \;,\; x_3 = \rho \, \sin \phi \;,
\eea
and introducing the coordinates
\bea
U = \tau - v \;\;\;,\;\;\; V = \tau +  v  \;\;\;,\;\;\; w =  \rho \, e^{i \phi} 
\;\;\;,\;\;\; {\bar w} =  \rho \, e^{-i \phi} ,
\eea
we obtain the line element \eqref{flat31},
\bea
 ds^2  = - d{\tau}^2 + dv^2 +  d\rho^2 + \rho^2 d\phi^2 =  - dU dV + dw d {\bar w} \;.
  \eea
Using
\bea
dx_0 \wedge dx_1 &=&  \frac12 dU \wedge d V \;, \nonumber\\
dx_2 \wedge dx_3 &=&
\rho \,  d \rho\, \wedge d \phi = \frac{i}{2} d w \wedge d {\bar w} \;, 
\eea
we obtain
\bea
 \F_{01} \, dx_0 \wedge dx_1 &=&  \frac12 \F_{01} \, dU \wedge d V = \F_{U V } \, dU \wedge d V \;, \nonumber\\
 \F_{23} \, dx_2 \wedge dx_3 &=& \frac{i}{2} \F_{23}\, d w \wedge d {\bar w} = 
\F_{w \bar w} \, d w \wedge d {\bar w} \;, 
\eea
and we infer the relations 
\bea
\F_{U V } = \frac12 \F_{01} \;\;\;,\;\;\; \F_{w \bar w} =  \frac{i}{2} \F_{23} \;.
\eea
Then, using \eqref{asdrel13}, we obtain
\bea
\F_{UV} =  \frac12 \F_{01} \ = -\frac{i}{2} \F_{23} = -\F_{w {\bar w}} \;.
\eea
Similarly, one infers the relations 
\bea
\F_{U \bar w} = \F_{V w} = 0 \;.
\eea
If we now interchange $U$ and $V$ we obtain
\bea
\F_{UV} = \F_{w {\bar w}} \;\;\;,\;\;\; \F_{U w} = \F_{V \bar w} = 0 \;.
\eea
Thus, by interchanging $U$ and $V$ the anti self-dual equations get mapped to the self-dual equations and vice-versa.

\section{From $2D$ to $4D$ \label{sec:2D4D}}

In Section \ref{sec:grYMm} we constructed the mapping of gravitational solutions to solutions of the self-dual Yang-Mills equations in four dimensions by starting from the latter equations and performing dimensional reductions of them to two dimensions. Here we present a different approach to this mapping, 
by starting on the gravitational side in two dimensions and expressing the 1-form $A$ in two dimensions as a 1-form in four space-time dimensions, as follows.

Let us consider a solution $M$ to the gravitational field equations \eqref{emotion2d}. The associated 
matrix 1-form $A = A_{\rho} d\rho + A_v dv$ satisfies the field equation \eqref{eoml1} as well as
$F_{\rho v} = 0$, c.f. \eqref{FAA}.
Next, we view $A$ as a 1-form in a four-dimensional flat space-time with 
lightcone coordinates $(U, V, X, Y)$,
\bea
A = A_U d U + A_V d V + A_X dX + A_Y d Y\,.
\label{A4dcomp}
\eea
The map between the lightcone coordinates 
$(U, V, X, Y)$ and the Weyl coordinates $(\rho,v)$ depends on the space-time signature. For example, for the case of signature $(1,3)$, it is given by
\eqref{UVwwb} and \eqref{wbwrt}.
The four components $(A_U, A_V, A_X, A_Y)$ are expressed in terms of $(A_{\rho}, A_v)$. Since the latter have to satisfy $F_{\rho v} =0$, this will impose conditions on the field strength components constructed from 
\eqref{A4dcomp}, which will be discussed below. 
Let us now assume that these four components are given in terms of 
first-order derivatives of two matrix functions
$\Psi (U,V, X, Y)$ and ${\tilde \Psi}(U, V, X, Y)$, 
\bea
A_U =  \partial_{Y} {\tilde \Psi} \;\;\;,\;\;\; A_V = 
- \partial_X \Psi\;\;\;,\;\;\;
A_X =  \partial_V {\tilde \Psi}\;\;\;,\;\;\; 
A_Y  = - \partial_U \Psi\;,
\label{Apotpsps}
\eea
and that $\Psi (U,V, X, Y)$ and ${\tilde \Psi}(U, V, X, Y)$ satisfy
\bea
\left( \partial_U \partial_V - \partial_X \partial_Y \right) {\Psi} + [\partial_U \Psi, \partial_X \Psi] = 0 \;,  \nonumber\\
\left( \partial_U \partial_V - \partial_X \partial_Y \right) {\tilde \Psi} 
+ [\partial_Y {\tilde \Psi}, \partial_V {\tilde \Psi}]  = 0 \;.
\label{harpsitpsi}
\eea
Now we consider the following two sets, $(0, A_V, 0, A_Y)$ and $(A_U, 0, A_X, 0)$. We observe that each of these two sets, when viewed by itself,
satisfies the self-duality conditions 
\bea
F_{UV} = F_{XY} \;\;\;,\;\;\; F_{UX} = 0 \;\;\;,\;\;\; F_{V Y } = 0 
\label{sdrelUVXY}
\eea
by virtue of \eqref{Apotpsps} and \eqref{harpsitpsi}.
For instance, if we consider the set $(0, A_V, 0, A_Y)$, then $F_{UX} = 0$ is trivially satisfied, since 
for this set $A_U = A_X = 0$. The condition $F_{V Y } = 0$ holds by virtue of the first equation in \eqref{harpsitpsi}, while the condition $F_{UV} = F_{XY}$ holds by virtue of \eqref{Apotpsps}.
Thus, there are two non-trivial conditions, which precisely equals the number of equations 
(namely \eqref{eoml1} and \eqref{FAA}) that $(A_{\rho}, A_v)$ have to satisfy. It is thus tempting to identify the condition $F_{V Y } = 0$ with $F_{\rho v} = 0$, and to identify 
the condition $F_{UV} = F_{XY}$ with the field equation \eqref{eoml1}. This turns out to be indeed the case in all the examples considered in this paper. 
Similar considerations hold for the set $(A_U, 0, A_X, 0)$. 

Since the two sets 
$(0, A_V, 0, A_Y)$ and $(A_U, 0, A_X, 0)$ are constructed from the same data (namely from $(A_{\rho}, A_v)$ and the coordinate transformation relating the lightcone coordinates $(U, V, X, Y)$ to the  coordinates used in the four-dimensional line element \eqref{4dWLP}), they carry the same information, and hence the matrix functions  $\Psi (U,V, X, Y)$ and ${\tilde \Psi}(U, V, X, Y)$ should be related.
This expectation has been borne out in all the examples studied in this paper. In these examples we find
that 
\bea
{\tilde \Psi} = e^{- 2 i \phi} \, \Psi \;.
\eea
We also note that in these examples $\Psi$ and $\tilde \Psi$ are complex conjugates of one another.

As an illustrative example, let us consider the Schwarzschild solution discussed in Section \ref{sec:exam12}. Using \eqref{Pschwarzm} and \eqref{PsiP}, we obtain the following expressions for $\Psi$ and $\tilde \Psi$ when expressed in terms of the lightcone coordinates given in \eqref{UVwwb} and \eqref{wbwrt},
\bea
\Psi &=& m \, \frac{V-U}{Y \, \sqrt{\frac14 (U-V)^2 + X Y }} \;,
\nonumber\\
{\tilde \Psi } &=& m \, \frac{ V-U}{X \, \sqrt{\frac14 (U-V)^2 + X Y }} \;.
\eea
Using \eqref{Apotpsps}, we obtain
\bea
A_U &=&  
4 m \, \frac{ U-V}{ \left( (U-V)^2 + 4 X Y \right)^{3/2} } \;, \nonumber\\
A_X &=&  
8 m \, \frac{ Y}{\left( (U-V)^2 + 4 X Y \right)^{3/2} }\;, \nonumber\\
A_V &=& 
4 m \, \frac{ V-U}{ \left( (U-V)^2 + 4 X Y \right)^{3/2} } \;, \nonumber\\
A_Y  &=& 
8 m \, \frac{ X}{\left( (U-V)^2 + 4 X Y \right)^{3/2} }\;.
\eea
It is straightforward to check that each of the two sets $(0, A_V, 0, A_Y)$ and $(A_U, 0, A_X, 0)$ satisfies the self-duality conditions \eqref{sdrelUVXY}. Both sets are related by interchanging $(U, X)$ with $(V, Y)$. We note that were we to pick 
any of the sets  $(A_U, 0, 0, A_Y)$ and $(0, A_V, A_X, 0)$, each of them would satisfy anti self-duality conditions, c.f. Appendix
\ref{sec:asdymds}.

Since the two sets $(0, A_V, 0, A_Y)$ and $(A_U, 0, A_X, 0)$ carry the same information,
we pick the set $(0, A_V, 0, A_Y)$
and declare it
to be the solution to the self-duality conditions \eqref{sdrelUVXY} that we associate to the gravitational solution encoded in $(A_{\rho}, A_v)$.

\providecommand{\href}[2]{#2}\begingroup\raggedright\endgroup

\end{document}